\documentclass[
reprint,
 amsmath,amssymb,
 aps,
pra,
]{revtex4-1}

\usepackage{graphicx}% Include figure files
\usepackage{dcolumn}% Align table columns on decimal point
\usepackage{bm}% bold math
\usepackage{hyperref}% add hypertext capabilities

\usepackage{braket} %introduced for Dirac notation
\usepackage[caption=false]{subfig}
\usepackage{mathrsfs} %introduced for 
\usepackage{physics} %introduced for \abs{ \mathscr{} characters}
\usepackage[retainorgcmds]{IEEEtrantools} %introduced for IEEEeqnarray environment
\usepackage{lipsum} %introduced for lorem ipsum text
\usepackage{amsthm} %introduced for proof environment
\usepackage{dsfont} %introduced for \mathds (to write the identity operator as \mathds 1)

\usepackage{}

\newcommand{\eqnref}[1]{Eq.~\ref{#1}}
\newcommand{\secref}[1]{Sec.~\ref{#1}}
\newcommand{\figref}[1]{Fig.~\ref{#1}}
\newcommand{\appref}[1]{Appendix~\ref{#1}}

\newlength{\argl}
\newlength{\argh}
\newlength{\tildel}
\newlength{\tildec}

\begin{document}

\preprint{APS/123-QED}

\title{Multi-Frequency Atom-Photon Interactions}

\author{Ben Yuen}
 \email{benjamin.yuen@physics.ox.ac.uk}
\affiliation{%
Clarendon Laboratory\\
Parks rd\\
Oxford\\
OX1 3PU
}%
\author{Christopher J. Foot}
\affiliation{%
Clarendon Laboratory\\
Parks rd\\
Oxford\\
OX1 3PU
}
%\author{Axel Kuhn}
%\affiliation{%
%Clarendon Laboratory\\
%Parks rd\\
%Oxford\\
%OX1 3PU
%}
 %\altaffiliation[Also at ]{Physics Department, XYZ University.}%Lines break automatically or can be forced with \\
%\author{Second Author}%

%\collaboration{MUSO Collaboration}%\noaffiliation

%\author{Charlie Author}
% \homepage{http://www.Second.institution.edu/~Charlie.Author}
%\affiliation{
% Second institution and/or address\\
% This line break forced% with \\
%}%
%\affiliation{
% Third institution, the second for Charlie Author
%}%
%\author{Delta Author}
%\affiliation{%
% Authors' institution and/or address\\
% This line break forced with \textbackslash\textbackslash
%}%
%
%\collaboration{CLEO Collaboration}%\noaffiliation

\date{\today}

\begin{abstract}
We present a formalism that enables the analytic calculation of the interaction of a spin-half particle with a polychromatic electromagnetic field. This powerful new approach provides a clear physical picture even for cases with highly degenerate energy levels, which are complicated to interpret in the standard dressed-atom picture.
Typically semi-classical methods are used for such problems (leading to equations that are solved by Floquet theory). Our formalism is derived from quantum electrodynamics and thus is more widely applicable. In particular it makes accessible the intermediate regime between quantum and semi-classical dynamics. We give examples of the application to multi-frequency multi-photon processes in strong fields by deriving the Hamiltonians of such systems, and also to the dynamics of weak fields at long times for which semi-classical methods are insufficient.
%\begin{description}
%\item[Usage]
%Secondary publications and information retrieval purposes.
%\item[PACS numbers]
%May be entered using the \verb+\pacs{#1}+ command.
%\item[Structure]
%You may use the \texttt{description} environment to structure your abstract;
%use the optional argument of the \verb+\item+ command to give the category of each item. 
%\end{description}
\end{abstract}

\pacs{Valid PACS appear here}% PACS, the Physics and Astronomy
                             % Classification Scheme.
%\keywords{Suggested keywords}%Use showkeys class option if keyword
                              %display desired
\maketitle

%\tableofcontents

The interaction of matter with electromagnetic fields gives exquisitely fine control over quantum systems. It is used, for example, to implement single-qubit operations on ions, neutral atoms and molecules, nuclear and electronic spins, and superconducting circuits~\cite{ladd2010}. These can be controlled by \emph{monochromatic} fields at radio, microwave and optical frequencies.
Yet we live in a \emph{polychromatic} world which is enriched by a far wider range of light-matter processes that we are only beginning to explore. 

Recently, new paradigms in multifrequency systems have emerged such as Floquet time crystals \cite{wilczek2012, else2016} and synthetic gauge fields \cite{goldman2014} in periodically driven systems - where an interaction at one frequency is modulated by one or more others. 
There are also important open problems concerning decoherence \cite{schlosshauer2005}, transport \cite{rebentrost2009}, and equilibration \cite{gogolin2016} which are intimately connected with a small quantum system coupled to a bath containing a large number of modes. 
There is a clear need for a comprehensive understanding of systems driven by multi-frequency fields and we present a powerful formalism for achieving this, in this paper.

Floquet's theorem is used in the mathematical description of systems that are periodic in time, and analogously Bloch's theorem is used for periodic structures in space.
Using Floquet's theorem, a quantum system driven by a periodic interaction can be described in terms of a basis of quasi-energy eigenstates. Originally, it was shown by Shirley \cite{shirley1965} that a two-level quantum system interacting with a classical, monochromatic field can be described by a Schr\"odinger equation with an effective Floquet Hamiltonian, near resonance, that is equivalent to the optical Bloch equations (in the absence of spontaneous emission).
%Floquet's theorem - a temporal analogue Bloch's theorem - provides a semi-classical analysis of multi-frequency systems of a periodic nature by extending the system on a basis of quasi-energies \cite{shirley1965}. Starting from a classical, periodic field which drives a quantum system, Floquet's theroem allows the decomposition of the system onto a basis of quasi-energy states. 
%Originally, Shirley applied Floquet's method to a two-level systems interacting with a monochromatic laser field. Shirley showed that the Schr\"odinger equation under the effective Floquet Hamiltonian near resonance \cite{shirley1965} is equivalent to the optical Bloch equations in the absence of spontaneous emission. 
Subsequently the Floquet approach has found a wide range of applications in quantum optics including periodically driven optical lattices \cite{holthaus2015}, quantum beam splitters \cite{wang2016}, and strong coupling of two-level atoms in periodic fields \cite{barata2000}. Methods typically involve a perturbative expansion in the extended Floquet space to find a suitable effective Hamiltonian when the modulation is either fast\cite{eckardt2015}, or slow\cite{novivcenko2017} compared to the other energy scales of the system. 

Quantum electrodynamics provides a complete picture of atom-photon interactions.
The Jaynes-Cummings model \cite{Jaynes1963, Shore1993}, describes the elementary case of a two-level system interacting with a second-quantised monochromatic field. In this model there is a ladder of energy levels corresponding to states of the combined atom-photon system which are called 'dressed states'.
%The simplest interaction of a two-level system with a second quantised monochromatic field is described by the Jaynes-Cumming model \cite{Jaynes1963, Shore1993}.%, and is almost synonymous with the `dressed-atom' picture of quantum optics.
%The Jaynes-Cummings model describes a ladder of atom-photon states which interact in pairs to form `dressed states'.
When the field is in a highly excited coherent-state the behaviour of these states tends towards a semi-classical picture where photon Fock states may be represented by Floquet quasienergy states. When the number of photons is small, or the field is in a nonclassical state, the behaviour is drastically different. Strikingly, Rabi oscillations rapidly collapse when the field is in a weakly excited coherent-state, with partial revivals at later times \cite{Eberly1980, Rempe1987}.

The dressed-atom picture \cite{Cohen94} - where the \emph{dressed} atom is surrounded by a cloud of self interactions mediated by the quantised electromagnetic field - has succeeded in describing a multitude of light-matter interactions over the past 50 years, from early work on radiofrequency transitions and optical pumping in atomic vapours \cite{Cohen94, Haroche1970} to ground breaking descriptions of laser cooling \cite{Dalibard1985} and cavity quantum electrodynamics \cite{Jaynes1963}.
%In quantum optics this picture contains three essential ingredients, the atom, the electromagnetic field and the interaction between them. 
Many of these effects have emerged from the extension of the Jaynes-Cummings model to include multi-level atoms.

Here we extend this approach to consider multi-frequency fields in a generalised framework which we apply to the Jaynes-Cummings and Rabi model.
In contrast to the single-frequency Jaynes-Cummings model, its multi-frequency relative is not equivalent to a set of closed two-level systems. This allows for a much wider variety of atom-photon processes to arise, but also makes multi-frequency atom-photon systems extremely difficult to solve exactly. Thus perturbative or other approximate methods must be employed.
Furthemore, approximate techniques - perturbative, numerical or otherwise - are also faced with a major challenge in that products of Fock states describing a multi-frequency state are energy degenerate when the frequencies are related by a rational number. This causes many approximate solutions to diverge. The degeneracy problem is worst when the frequencies are low harmonics, such as found in some rf systems, or in optical fields with harmonic spectra.

Motivated by the challenge of reducing the complexity of multi-frequency interactions and the degeneracy problem, we introduce a non-degenerate formalism for the field when each mode can be described by a coherent-state, or more generally a coherent-state representation of phase space \cite{glauber1963}. Here we investigate the quantum electrodynamics in strong fields using dressed-atom techniques generalised to multiple frequencies, and in weak fields where semi-classical approaches such as the Floquet theory break down. Remarkably, we find our non-degenerate theory of the field produces accurate results even for very low photon numbers when the spin-field coupling is strong compared to the frequency difference between field modes. In other work we have applied this formalism to investigate ultracold-atoms dressed by multiple radio frequencies \cite{harte2018, bentine2017}, and used it to understand transitions between multi-frequency dressed states \cite{luksch2018}. In \cite{yuen2018analytic} we use this formalism to derive accurate analytic expressions for the time evolution operator for a spin-half system in a multi-frequency coherent field.

We begin in \secref{subsec:fockbasisproblem} by highlighting the challenges that aries when using the conventional Fock basis for describing an atom dressed by a multi-frequency field. Section \ref{sec:basis} introduces an alternative, non-degenerate basis, and we derive how the field operators act on these non-degenerate basis states. To demonstrate the applicability of this basis we perform numerical (\secref{sec:calculations}) and perturbative calculations (\secref{sec:effectiveHs}) for the eigenenergies and dynamics of an atom in a multi-frequency field. In \secref{sec:quantumclassicalboundary} we consider the weak field limit and the limitations of this new formalism; How it breaks down sheds light on the boundary between semi-classical and quantum behaviour. We conclude in \secref{sec:conclusion}.

\section{The Polychromatic Jaynes-Cummings Model in the Fock basis} \label{subsec:fockbasisproblem}

\begin{figure*}[ht]
	\subfloat[Degenerate basis \label{fig:basisdiagram}]{
			\begin{picture}(280,250)
				\includegraphics[width=280pt]{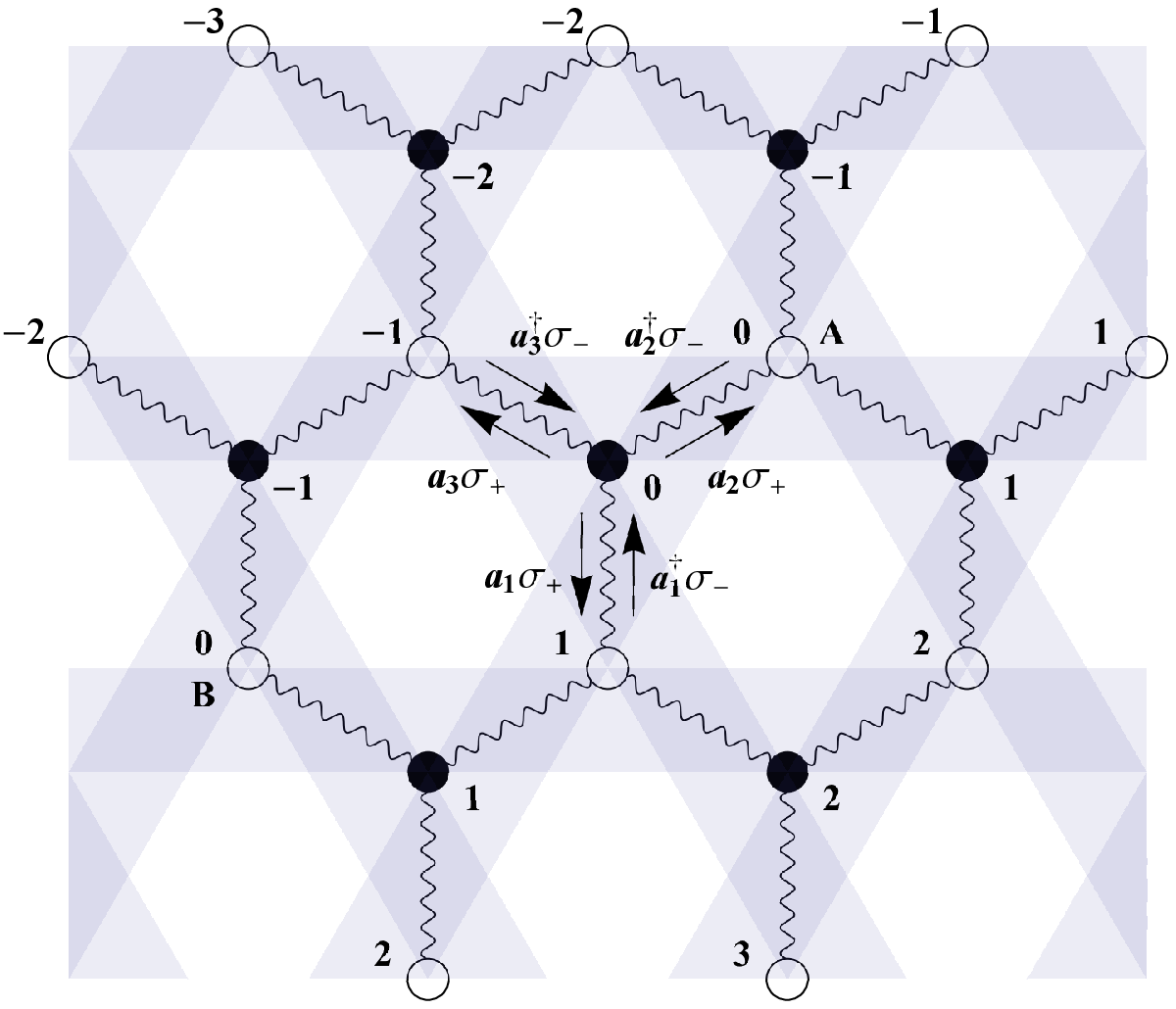}
			\end{picture}
	}
	\subfloat[Energy levels \label{fig:energylevels}]{
			\begin{picture}(210,250)
				\put(100,5){\includegraphics[width=110pt]{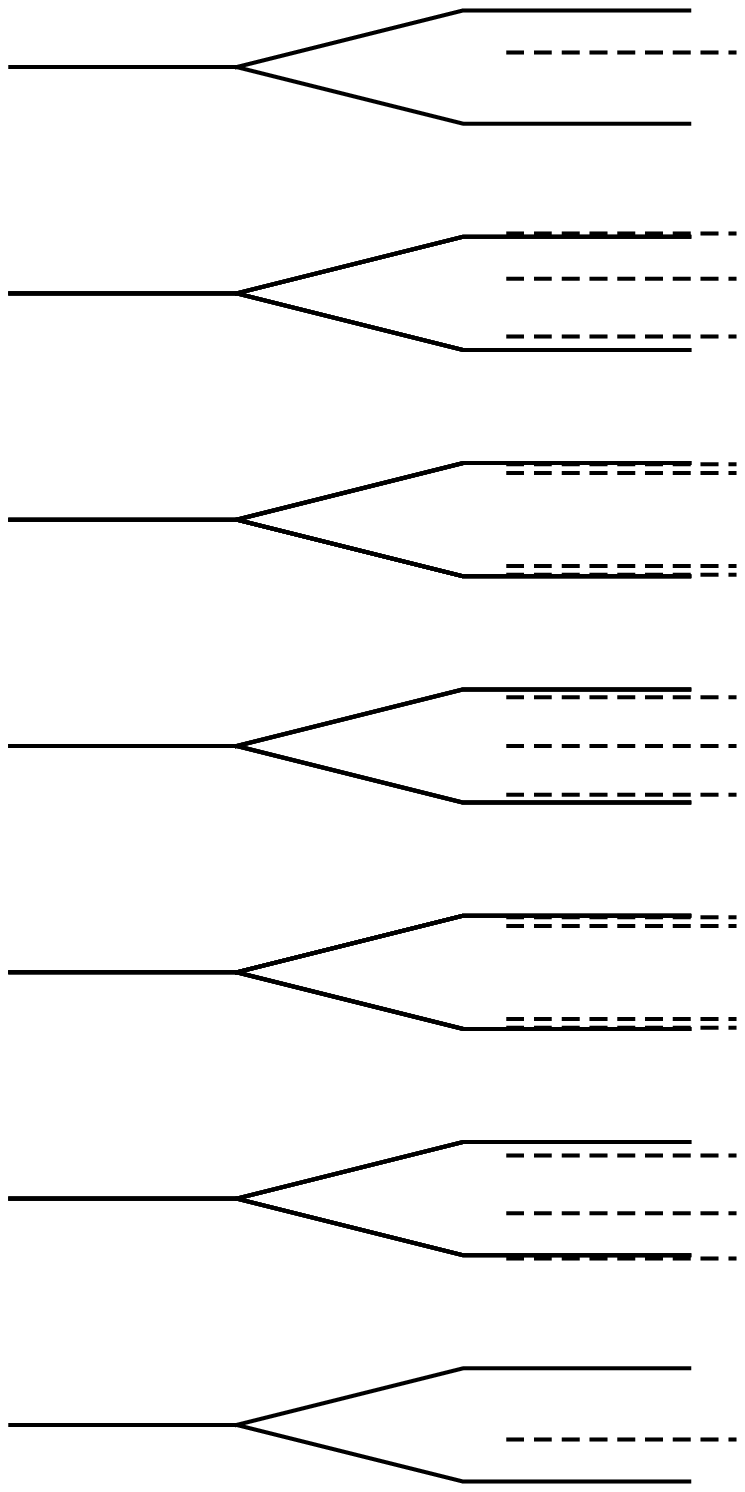}}
				\put(57,12){$\ket{10-2\uparrow}$}
				\put(4,41){$\ket{10-1\downarrow}$}
				\put(45,41){$\ket{1-1-1\uparrow}$}
				\put(56,53){$\ket{01-2\uparrow}$}
				\put(15,76){$\ket{1-10\downarrow}$}
				\put(56,76){$\ket{1-20\uparrow}$}
				\put(15,88){$\ket{01-1\downarrow}$}
				\put(56,88){$\ket{00-1\uparrow}$}
				\put(68,110){$\ket{000\downarrow}$}
				\put(27,110){$\ket{0-10\uparrow}$}
				\put(50,122){$\ket{-11-1\uparrow}$}
				\put(23,146){$\ket{-100\uparrow}$}
				\put(60,146){$\ket{-110\downarrow}$}
				\put(15,157){$\ket{0-11\downarrow}$}
				\put(56,157){$\ket{0-21\uparrow}$}
				\put(60,180){$\ket{-101\downarrow}$}
				\put(12,180){$\ket{-1-11\uparrow}$}
				\put(60,192){$\ket{-210\uparrow}$}
				\put(60,220){$\ket{-201\uparrow}$}
			\end{picture}
	}
\caption{\label{fig:degeneratebasisdiagram} Feynman diagram (a) and dressed state energies (b) of a two-level atom interacting with a three frequency electromagnetic field. The Feynman diagram, which illustrates states with different spin and photon number (vertexes) and their interactions (wavy lines), demonstrates the complexity of the multifrequency Jaynes-Cummings model in the Fock-spin basis. The energy level diagram highlights the erroneous calculation of the dressed energies in this basis (dashed lines) in contrast with the correct result (solid lines). 
The field frequencies are $j \omega_f$ and $(j\pm1) \omega_f$, and the energy levels are shown when the spin is resonant with $\omega_0=j\omega_f$.
In (a) the open (filled) circles are spin up (down) states with $N+\Delta N$, each labelled by their energy difference from the state in the centre of the diagram. The orientation of the wavy lines indicates the mode of interaction, and the shaded bands contain states of constant photon number in mode 1, 2 or 3. States such as $A$ and $B$ are degenerate, which cause problems in perturbative or numerical calculations of the eigenstates of \eqnref{eq:polychromaticJCM}. This can be seen in (b) by the incorrect appearance of singlets and triplets, and the disagreement of the level shifts. The states in (b) are labelled with shorthand $\ket{\Delta n_1, \Delta n_2, \Delta_n , m}$ for $\ket{n_1+\Delta n_1, n_2 + \Delta n_2, n_3 + \Delta n_3, m}$ such that $\ket{000 m} = \ket{n_1,n_2,n_3 m}$.
}
\end{figure*}

We consider a two-level system, such as an atom, interacting with a multi-frequency field with fundamental frequency $\omega_f$ \footnote{The frequency $\omega_f$ need only be the greatest common denominator of the frequencies under consideration, but in principle could be the fundamental frequency $2 \pi c/L$ when one imposes periodic boundary conditions to quantise the electromagnetic field. Thus, requiring that the field frequencies are rationally related in this model is not an overly restrictive assumption.},
%The atom-photon interaction we shall consider in this section, \secref{sec:calculations} and \secref{sec:effectiveHs} are 
described by the polychromatic Jaynes-Cummings model with Hamiltonian
\begin{equation} \label{eq:polychromaticJCM}
	H = \tfrac12 \omega_0 \sigma_z + \sum_i \left[k_i \omega_f a_{k_i}^{\dagger} a_{k_i} + g_{k_i} \left( a_{k_i} \sigma_+ + a_{k_i}^{\dagger} \sigma_-\right) \right].
\end{equation}
The rotating-wave approximation has been made and this form is relevant for the systems considered in \secref{sec:calculations} and \secref{sec:effectiveHs}; the Rabi model without this approximation is treated in \secref{sec:quantumclassicalboundary}.
%In \secref{sec:quantumclassicalboundary} we will also consider the Rabi model - where the rotating wave approximation has not been made.
In this model $\pm \tfrac12 \omega_0$ are the spin eigenenergies, the field modes are indexed by $i$ and the $i^{\mathrm{th}}$ mode has frequency $k_i \omega_f$ where $k_i$ are integers and need not be commensurate. $g_{k_i}$ are the coupling spin-field constants. More generally the mode frequencies do not need to be commensurate.

A natural basis to describe the coupled atoms and photons comprises of Fock-spin states $\ket{n_{k_1},n_{k_2},... ,m}$, where $n_{k_i}$ is the number of photons in mode $k_i$ and $m \hbar$ the atom's spin projection. These are eigenstates of $H_0 = \tfrac12 \omega_0 \sigma_z + \sum_{\{k \}} k \omega_f a_{k}^{\dagger} a_{k}$ and an extension of the basis $\{ \ket{n,m} \vert n \in \mathds Z^+ , m=\pm\tfrac12 \}$ used extensively for the single-frequency Jaynes-Cummings model. However, calculations performed in this basis are often divergent. The problems associated with this basis for the multi-frequency model are caused by (\emph{i}) the nature of multi-frequency interaction $V= \sum_{\{k \}} g_k \left( a_k \sigma_+ + a_k^{\dagger} \sigma_- \right)$ and (\emph{ii}) degeneracy of these states. In section \ref{sec:basis} we introduce a new non-degenerate basis upon which the dynamics of the multi-frequency model \eqnref{eq:polychromaticJCM} can be solved using a wide range of conventional methods. In the remainder of this section we elaborate on the key differences between the single and multi-frequency Jaynes-Cummings model and the problem of working with Fock-spin states.

In striking contrast to the single-frequency model is that a vast number of different states can now interact with each other. For a single-frequency field only pairs of states $\ket{n,-\tfrac12}$ and $\ket{n-1,\tfrac12}$ are permitted to interact - under the rotating wave approximation - by the terms $a \sigma_+$ and $a^{\dagger} \sigma_-$. For multiple frequencies, atoms can climb the Jaynes-Cummings ladder by interacting on alternate modes.
The extended range of interactions, together with energy degeneracies of the Fock states $\ket{n_1,n_2,...}$ cause the standard numerical or perturbative approaches to breakdown due to divergences in their solutions.

It is well known for example, that an energy degeneracy between a pair of interacting states $\ket{\psi_1}$ and $\ket{\psi_2}$ causes the perturbative solutions to the Schr\"odinger equation to diverge due to the energy denominator $E_1-E_2=0$. Such degeneracies are common for multi-frequency fields of \eqnref{eq:polychromaticJCM}. Consider for example a field with three commensurate frequencies (i.e. $k_1$, $k_2=k_1+1$ and $k_3=k_1+2$), and the interaction between states $\ket{n_1,n_2,n_3,m}$ and $\ket{n_1+1,n_2-2,n_3+1,m}$. These states are degenerate since $k_1-2k_2+k_3 = 0$. They also interact via a fourth order interaction $\propto a_{k_1}^{\dagger} \sigma_- a_{k_2} \sigma_+ a_{k_3}^{\dagger} \sigma_- a_{k_2} \sigma_+$. The energy shift caused by this interaction incorrectly appears to diverge when calculated perturbatively. So called `non perturbative' analytic methods such as the resolvent formalism employed in \secref{sec:effectiveHs} also suffer from such degeneracies for the same reason.

Simple numerical approaches also suffer from these degeneracies.
To diagonalise \eqnref{eq:polychromaticJCM} numerically we must truncate the basis over a finite range of values $n_k$ for each mode. An efficient truncation method starts with state $\ket{n_1,n_2,n_3,-\tfrac12}$ for a three mode field for example, and includes any states which can be connected by applying $a_k \sigma_+$ and $a_k^{\dagger} \sigma_-$ up to some maximum number $N$ times. For example, including up to $N=1$ interactions, the basis consists of $\ket{n_1,n_2,n_3,-\tfrac12}$ together with $\ket{n_1-1, n_2, n_3,\tfrac12}, \ \ket{n_1, n_2-1
, n_3,\tfrac12}$ and $\ket{n_1, n_2, n_3-1,\tfrac12}$. For $N=2$ our basis includes six further states connected to $\ket{n_{k_1}, n_{k_2}, n_{k_3},-\tfrac12}$ via $a_{k_j}^{\dagger} \sigma_- a_{k_i} \sigma_+$ for all combinations $i\neq j$. The truncated basis is shown diagrammatically in \figref{fig:basisdiagram} for $N=3$ along with their energies. 
Even in this limited basis degeneracies are abundant; There are seven pairs of states which have the same energy and spin.

The problem encountered numerically is that the basis must be truncated and subsequently the degenerate states are not all surrounded by a similar set of interactions. Look for example at the state spin up state $\ket{n_1,n_2-1,n_3,\tfrac12}$ with energy $0$ which lies near the centre of the basis in \figref{fig:basisdiagram} and is marked $A$. Its degenerate partner $\ket{n_1-1,n_2+1,n_3-1,\tfrac12}$ , marked $B$ lies on the left side of this figure. The energy of $\ket{n_1,n_2-1,n_3,\tfrac12}$ is predominantly shifted by the interactions with the three neighbouring states, while $\ket{n_1-1,n_2+1,n_3-1,\tfrac12}$ only interacts directly with two neighbouring state in this truncated basis. Thus, the dressed energies of these states, which ought to be degenerate, are not. This can be seen by the dressed energies shown in \figref{fig:energylevels}. This problem cannot be solved simply by expanding the basis as further degenerate states will also near the edge of the extended basis - extending the basis eventually leads to bands of dressed energies with erroneous intra-band avoided crossings.

The solution to the problems caused by degenerate multi-frequency Fock states is to describe the field using sets of non-degenerate states $ \{ \ket{N} \}$ which we derive below. Each state is labelled by $N$ and has energy $N \omega_f$. Each set spans a non-degenerate subspace of Fock space which is sufficient to describe a multimode field in a product of coherent-states. We derive the multi-frequency number operator \eqnref{eq:diagfieldenergy}, and show that the action of the field creation and annihilation operators are $a_k \ket{N} = \alpha_k (\gamma_{N-k}/\gamma_N) \ket{N-k}$ and $a_k^{\dagger} \ket{N} = a_k^* (\gamma_N/\gamma_{N+k}) \ket{N+k}$ (\eqnref{eq:offdiagmatrixelements}).
These relations can be understood from the conservation of energy since $a_k$ annihilates a photon with energy $k \omega_k$.
We define $\gamma_N$ in \eqnref{eq:defgammaN} below, and show that in the limit of large photon numbers $\gamma_N/\gamma_{N\pm k} \approx 1$ using \eqnref{eq:gammaapprox}. 
With these relations one can study the quantum electrodynamics of multi-frequency atom photon interactions in coherent fields using the non-degenerate basis $\{ N \}$. 
The matrix elements of the Jaynes-Cummings Hamiltonian for example are readily evaluated in this basis using \eqnref{eq:approxoffdiagmatrixelements} and \eqnref{eq:diagonalmatrixelements}.
We explore these dynamics for strong fields in \secref{sec:effectiveHs}, and weak fields in \secref{sec:quantumclassicalboundary}.

\section{A non-degenerate dressed basis for a spin-half particle in a multi-frequency field} \label{sec:basis}

In this section we define the non-degenerate basis and derive the action of the field operators in this basis.
We assume each frequency mode of the field is in a coherent-state
\begin{equation}
	\ket{\alpha_k} = e^{- \abs{\alpha_k}^2 / 2} \sum_{n_k=0}^{\infty} \frac{\alpha_k^{n_k}}{\sqrt{n_k!}} \ket{n_k}
\end{equation}
Thus, the field is in the tensor product of coherent-states defined by the set of complex numbers $\alpha_k$, which we write as
\begin{equation}
	\ket{ \{ \alpha_k \}} = {\prod_k}^{\otimes} \ket{\alpha_k}.
\end{equation}
We next express such field states in a non-degenerate basis $\{\ket{N} \}$.

The Hilbert space of the field $\mathscr H_F$ is spanned by the set of Fock states $\{ \ket{n_k} \}$. We partition this space into distinct subspaces $\mathscr E_N$ of energy $N \omega_f$ (see \appref{ap:partition} for proof). Formally, $\mathscr E_N = \left \{ \ket{\psi} \vert H_F\ket{\psi} = N \hbar \omega_f \ket{\psi} \right \}$, where $N$ can be any positive integer.
The Hamiltonian $H_F = \sum_k k \hbar \omega_f a_k^{\dagger} a_k$ is the energy of the field minus the vacuum energy.

Let $P_N:\mathscr H_F \rightarrow \mathscr E_N$ be the projector onto the $N^{\text{th}}$ subspace. Given the state $\ket{\{\alpha_k \}}$ we define the state $\ket{N}$ as
\begin{equation}
	\ket{N} = \frac{P_N \ket{\{ \alpha_k \}}}{\sqrt{\bra{\{\alpha_k \}} P_N \ket{\{ \alpha_k\} } }},
\end{equation}
which has norm $\braket{ N } = 1$ since $P_N^2=P_N$. 

Since $\{ \mathscr E_N \}$ is a partition of $\mathscr H_F$, we can write the identity operator $\mathds 1 = \sum_{N=0}^{\infty} P_N$. Applying this to coherent-state $\ket{\{ \alpha_k \}}$ expands the field's state in terms of states $\ket{N}$,
\begin{equation} \label{eq:Nexpansion}
	\ket{ \{ \alpha_k\} } = \sum_{N=0}^{\infty} \gamma_N \ket{N}
\end{equation}
where 
\begin{equation} \label{eq:defgammaN}
\gamma_N = \sqrt{\bra{\{\alpha_k \}} P_N \ket{\{ \alpha_k\}}} \in \mathds R.
\end{equation} 
Importantly, this expansion of $\ket{\{ \alpha_k\}}$ is on a non-degenerate basis, unlike its expansion in terms of Fock state products. We note that the basis $\{ \ket{N} \}$ is specific to the set $\{ \alpha_k \}$ of complex numbers which describe the coherent-state of the field. Furthermore, the basis does not span $\mathscr H_F$ but is sufficient to describe this specific state of the field. This subspace $\{\ket{N} \} \subset \mathscr H_F$ is closed under the action of $a_j$, a property which is inherited from the coherent-state of the field. Although it is not strictly closed under $a_j^{\dagger}$, it approximately closes to the degree of $\braket{\alpha,1}{\alpha}$, the over-lap between a coherent-state and a photon added coherent-state \cite{agarwal1991}, which falls off rapidly for $\alpha$ larger than unity.

We now turn to the matrix elements of \eqnref{eq:polychromaticJCM} (and \eqnref{eq:shrodrabi}) in this basis. The field energy in the diagonal terms is 
\begin{equation} \label{eq:diagfieldenergy}
	\bra{N} \sum_k k \hbar \omega_f a_k^{\dagger} a_k \ket{N}=N \hbar \omega_f,
\end{equation}
since $\ket{N} \in \mathscr E_N$. For the off-diagonal terms we need to evaluate $\bra{N} a_j \ket{M}$ and its Hermitian conjugate $\bra{M} a_j^{\dagger} \ket{N}$. To do this we first prove that $a_j \ket{M} \in \mathscr E_{M-j}$.
Applying $H_F$ to $a_j \ket{M}$m,
	\begin{IEEEeqnarray*}{rCl}
		H_F a_j \ket{N} 
			&=& \hbar \omega_f \sum_{l=1}^{\infty} l a_l^{\dagger} a_l a_j \ket{N} \\
			&=& \hbar \omega_f \left[j a_j^{\dagger} a_j a_j+ a_j \sum_{l \neq j} l a_l^{\dagger} a_l \right] \ket{N} \\
			&=& \hbar \omega_f \left[j \left (a_j a_j^{\dagger} - 1\right ) a_j + a_j \sum_{l \neq j} l a_l^{\dagger} a_l \right] \ket{N} \\
			&=& \hbar \omega_f a_j \left[ -j + \sum_{l=1}^{\infty} l a_l^{\dagger} a_l \right] \ket{N} \\
			&=& (N-j) \hbar \omega_f a_j \ket{N}.
	\end{IEEEeqnarray*}
This shows that the state $a_j \ket{N}$ is an eigenstate of $H_F$ with energy $(N-j)\hbar \omega_f$. Thus $a_j \ket{N} \in \mathscr E_{N-j}$.

From this we note that $\bra{N} a_j \ket{M} = 0 \forall M \neq N+j$. We next evaluate $\bra{N} a_j \ket{\{ \alpha_k \}}$ in two ways. On the one hand,
\begin{IEEEeqnarray*}{rCl}
	\bra{N} a_j \ket{\{ \alpha_k \}} 
		&=& \bra{N} \alpha_j \ket{\{ \alpha_k \}} \\
		&=& \alpha_j \bra{N} \sum_M \gamma_M \ket{M} \\
		&=& \alpha_j \gamma_N. \IEEEyesnumber \label{eq:Najalphak1}
\end{IEEEeqnarray*}
On the other hand,
\begin{IEEEeqnarray*}{rCl}
	\bra{N} a_j \ket{\{ \alpha_k \} }
		&=& \bra{N} a_j \sum_M \gamma_M \ket{N} \\
		&=& \gamma_{N+j} \bra{N} a_j \ket{N+j} \IEEEyesnumber \label{eq:Najalphak2}
\end{IEEEeqnarray*}
since only $a_j \ket{M} = a_j \ket{N+j}$ is in the same subspace $\mathscr E_N$ as $\ket{N}$. Equating the two expression \eqnref{eq:Najalphak1} and \eqnref{eq:Najalphak2} then rearranging, we find
\begin{IEEEeqnarray*}{rCl} \IEEEyesnumber \label{eq:offdiagmatrixelements}
	\bra{N} a_j \ket{M} &=& \frac{\gamma_N}{\gamma_{N+j}} \alpha_j \delta_{M,N+j}. \IEEEyessubnumber
\end{IEEEeqnarray*}
Taking the Hermitian conjugate and switching labels $N$ and $M$,
\begin{IEEEeqnarray*}{rCl}
	\bra{N} a_j^{\dagger} \ket{M} &=& \frac{\gamma_{N-j}}{\gamma_N} \alpha_j^* \delta_{M,N-j}.\IEEEyessubnumber
\end{IEEEeqnarray*}
To evaluate these expression we must first evaluate $\gamma_N^2$.

The probability distribution of states $\ket{N}$ is given by $\bra{\{\alpha_k \}} P_N \ket{ \{ \alpha_k \} } = \gamma_N^2$. This distribution has mean value $\bar N = \sum_k k \abs{\alpha_k}^2$ since $\langle \hat N\rangle = \sum_k \langle  k a_k^{\dagger} a_k \rangle = \sum_k k \abs{\alpha}_k^2$, where $\langle \hat O \rangle$ is the expectation value of an operator $\hat O$. The variance of $\gamma_N^2$ is $\sigma_N^2 = \sum_k k^2 \abs{\alpha_k}^2$ using $\langle N^2 \rangle - \bar N^2$ and the commutator relation $[a_k,a_k^{\dagger}]=1$.
We show in \appref{ap:gammaapprox} that the coefficients $\gamma_N^2$ tend towards the Gaussian
\begin{equation} \label{eq:gammaapprox}
	\gamma_N^2 \approx \frac{k_0}{\sqrt{2 \pi} \sigma_N} \exp \left[ - \frac{(N-\bar N)^2}{2 \sigma_N^2} \right],
\end{equation}
in the limit that the excitation of each mode $\abs{\alpha_k}^2$ is large, where $k_0$ is the greatest common denominator of $\{k\}$.

From the distribution \eqnref{eq:gammaapprox} we see that the states with non-negligible population are those distributed within $\sim \sigma_N$ of $\bar N$. The value of $\sigma_N$ is bounded such that $\sigma_N \leq \sqrt {\bar N} \sum_k k$ since $\lambda_k \leq 1/k$. Thus the standard deviation $\sigma_N \ll \bar N$ provided that $\bar N \gg 1$ and $\sum_k k \ll \sqrt{\bar N}$. In this case, we see from \eqnref{eq:gammaapprox} that the ratio $\gamma_N/\gamma_{N+j}\approx 1$. Under this approximation the off diagonal matrix elements ($N \neq M$) of \eqnref{eq:polychromaticJCM} are
\begin{IEEEeqnarray*}{rCl} \IEEEyesnumber \label{eq:approxoffdiagmatrixelements}
	\bra{N,1/2} H \ket{M,-1/2} &=& \frac{g}{\sqrt 2}\alpha_j e^{i \mathbf k_j \cdot \mathbf r}\delta_{M,N+j},\IEEEyessubnumber \\
	\bra{N,-1/2} H \ket{M,1/2} &=& \frac{g}{\sqrt 2} \alpha_j^* e^{i \mathbf k_j \cdot \mathbf r}  \delta_{M,N-j}.\IEEEyessubnumber
\end{IEEEeqnarray*}
The diagonal terms are given exactly by
\begin{equation} \label{eq:diagonalmatrixelements}
	\bra{N,\pm 1/2} H \ket{N,\pm 1/2} = \pm \frac{1}{2}\hbar \omega_0 + N \hbar \omega_f.
\end{equation}

\section{Eigenenergies calculated in the non-degenerate basis} \label{sec:calculations}

\begin{figure}[!t]
%	\begin{tabular}{lr}
	\subfloat[Nondegenerate basis \label{fig:ndbasis}]{
%	\begin{subfigure}{0.58\columnwidth}
	\begin{picture}(150,300)
		\put(40,10){\includegraphics[width=50pt]{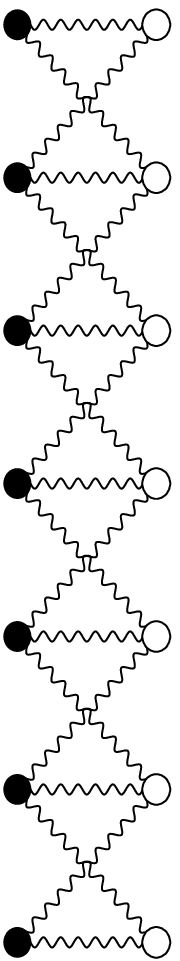}}
		\put(0,13){$\ket{N-3,\downarrow}$}
		\put(95,13){$\ket{N-3-k_2,\uparrow}$}
		\put(0,58){$\ket{N-2,\downarrow}$}
		\put(95,58){$\ket{N-2-k_2,\uparrow}$}
		\put(0,103){$\ket{N-1,\downarrow}$}
		\put(95,103){$\ket{N-1-k_2,\uparrow}$}
		\put(0,148){$\ket{N,\downarrow}$}
		\put(95,148){$\ket{N-k_2,\uparrow}$}
		\put(0,193){$\ket{N+1,\downarrow}$}
		\put(95,193){$\ket{N+1-k_2,\uparrow}$}
		\put(0,237){$\ket{N+2,\downarrow}$}
		\put(95,237){$\ket{N+2-k_2,\uparrow}$}
		\put(0,283){$\ket{N+3,\downarrow}$}
		\put(95,283){$\ket{N+3-k_2,\uparrow}$}
	\end{picture}
	}
	\subfloat[Energy levels \label{fig:ndlevels}]{
	\begin{picture}(69,295)
		\put(0,0){\includegraphics[width=69pt]{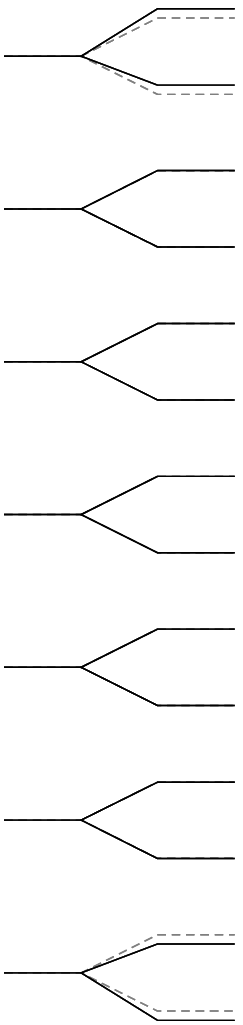}}
	\end{picture}
	}
\caption{\label{fig:ndbasisandlevels}
Feynman diagram (a) of a two-level atom in a three frequency field represented in the non-degenerate basis, and (b) the ladder of eigenenergies calculated using this basis. The Feynman diagram here demonstrates the simplicity of working in this basis, compared to \figref{fig:basisdiagram} which represents the same problem in the Fock-spin basis. Furthermore, the eigenenergies calculated in a truncated basis correctly form a set of doublet states, unlike those in \figref{fig:energylevels}. In (a) atom-photon states are represented by circles at each vertex as labelled, and the diagram is truncated after states that interact with $\ket{N, \downarrow}$ or $\ket{N-j,\uparrow}$ via a minimum of three photons.
The wavy lines show the interactions between them through the field with frequencies $\omega_1, \ \omega_2$ and $\omega_3$ equal to $(j-1) \omega_f, \ j \omega_f$ and $(j+1)\omega_f$. The state energy increases going up the page with resonant states shown at the same height. In this example $\omega_2$ is chosen to be resonant with the spin-half.
Thus, wavy horizontal lines represent interactions $\propto a_2 \sigma_+$ or $a_2^{\dagger} \sigma_-$. Diagonal wavy lines in one direction represent $a_1 \sigma_+$ and $a_1^{\dagger} \sigma_-$, and the other $a_3 \sigma_+$ and $a_3^{\dagger} \sigma_-$. At this resonance the energies shown in (b) of pairs of states $\ket{N, \downarrow}$ and $\ket{N-{k_2}, \uparrow}$ etc. are equal and form singlets. These are split into doublets in the dressed state basis as is typical in the Jaynes-Cummings model. Solid lines show state energies calculated in this truncated non-degenerate basis larger basis and are in very close agreement with more precise calculations in a large truncated basis which includes states up to 5 interactions from $\ket{N, \downarrow}$.
}
\end{figure}

Immediately one finds a multi-frequency model's Hamiltonian matrix can be diagonalised without the problems caused by level degeneracies.
The set of basis states can be efficiently truncated in the same way that produced the basis in \figref{fig:basisdiagram}. 
We consider again the three mode field with frequencies $(j-1)\omega_f, \ j \omega_f$ and $(j+1)\omega_f$ for some integer $j$. These modes are the same as in \secref{subsec:fockbasisproblem} when $j=1$.
Starting from a state $\ket{N,-\tfrac12}$ we include all states connected to this by up to $N$ interactions.

Figure \ref{fig:ndbasis} shows the diagram for $N=3$, which is significantly simpler than the degenerate diagram in \figref{fig:basisdiagram}. In fact the two diagrams become topologically equivalent if one identifies the degenerate states with one another.
Figure \ref{fig:ndlevels} shows some of the eigenenergies calculated in the non-degenerate basis truncated at $N=5$. In this case the energies do not suffer the problems seen in the degenerate basis. On the left hand side of this diagram lie the energy levels of pairs of states $\ket{n,-\tfrac12}$ and $\ket{n-j,\tfrac12}$, $\ket{n+1,-\tfrac12}$ and $\ket{n-j+1,\tfrac12}$, ... Each pair are resonant since we have taken $\omega_0-j \omega_f=0$. These dressed state energies (far right of \figref{fig:ndlevels}) are shifted apart by the interaction between them forming the doublets. The level shifts are calculated most accurately for states near the centre of the basis. When $N=3$ states near the edge of the basis (the highest or lowest energies in this example) can be seen to deviate from more accurate calculations with $N=5$.
\begin{figure}
	\includegraphics[height = 5.8cm]{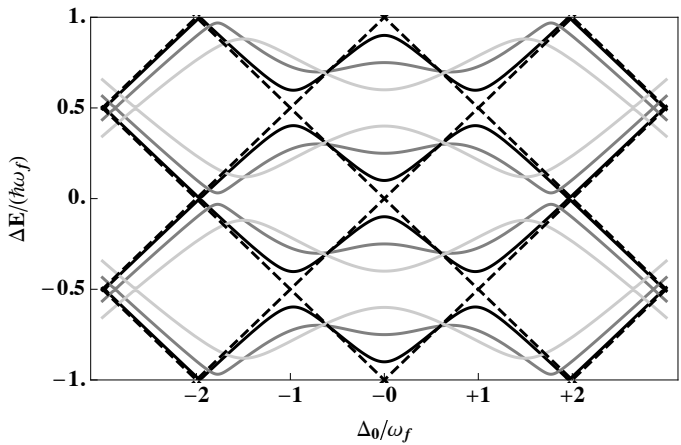}
\caption[]{\label{fig:evals3f} 
Energy levels of a spin-half in a three frequency field versus detuning between the spin-half and the mean field frequency. Each field is resonant at either $\Delta_0 = -1, \ \Delta_0 = 0$ or $\Delta_0 = 1$.
The dashed lines show energies $N\omega_f \pm \tfrac12 \Delta_0$ of the states $\ket{N, \pm \tfrac12}$. The interactions between these states causes avoided crossings in the dressed state energies (solid lines) at resonance $\Delta_0=0$ and $\pm1$. The Rabi frequencies are equal between the fields; $\Omega = 0.2 \omega_f, \ 0.5 \omega_f$ and $0.8 \omega_f$ for the black, grey and light grey lines.
The avoided crossings are shifted towards $\Delta_0$ due multi-frequency processes involving the off resonant fields. At $\Delta_0=\pm2$ further avoided crosses emerge due to three photon interactions, clearly visible for stronger Rabi frequencies. The light grey line shows how the the central three avoided crossings merge to form a single resonance as the Rabi frequencies approaches the field's frequency spacing $\omega_f$.
}
\end{figure}

Figure \ref{fig:evals3f} shows a generalisation of this energy level diagram, plotted as a function of $\Delta_0 = \omega_0-j \omega_f$ along the horizontal axis. The nondegenerate basis states (dashed lines) $\ket{N,\tfrac12}$ increase linearly with $\Delta_0$ while $\ket{N,-\tfrac12}$ decrease linearly. These basis states cross whenever $\Delta_0/\omega_f$ takes an integer value. We label each crossing by this integer value, $q$. For $q=0$ the states $\ket{N,-\tfrac12}$ and $\ket{N+j,\tfrac12}$ are resonant as they are in \figref{fig:ndbasisandlevels}. Generally the $q^{\mathrm{th}}$ resonance is between $\ket{N-\tfrac12}$ and $\ket{N-(j+q),\tfrac12}$.

The dressed levels (solid lines) are quite clearly forced apart for $q=-1,0,1$ predominantly due to the first order interactions with modes $j-1, \ j$ and $j+1$ respectively.
The $q=\pm1$ resonances are also shifted inwards slightly due to the level shift caused by second order interactions involving the off resonant modes. These shifts are the multi-frequency equivalent to the Bloch-Siegert shift. The contributing shifts to the $q=0$ resonance cancel when the amplitudes of modes $j\pm1$ are balanced.
Less prominent avoided crossings occur for $q=\pm2$. These become noticeable as the field amplitudes increase (grey solid line) and are caused by resonant three photon interactions. Using our non-degenerate basis we are able to investigate these multi-frequency resonances and level shifts analytically in the section below.

\section{Multi-photon processes of a spin-half particle in strong multi-chromatic field} \label{sec:effectiveHs}

We now derive effective Hamiltonians which describe the system near each of the crossings of the levels of $H_0$. For this we use the resolvent formalism, allowing us to account for the level shifts and interaction processes which couple these levels to an arbitrary precision.

Briefly, we introduce the formalism of the resolvent $G(z)=(z-H)^{-1}$ which is the advanced Green's function for the Shro\"odinger equation. Its Fourier transform is the time evolution operator and we can interpret $G(z)$ as the propagator in the complex frequency ($z$) space.
We are able to describe the dynamics between two energy states of $H_0$ (close to where they cross) using the projection of the resolvent $PG(z)P=(z-H_{\mathrm{eff}})^{-1}$ onto subspace $\mathscr E$ these states span. $H_{\mathrm {eff}}$ is the effective Hamiltonian between the two states and is given by
\begin{IEEEeqnarray}{rCl} \label{eq:Heffgeneral}
	H_{\mathrm{eff}} &=& P H P + PR(z)P 
\end{IEEEeqnarray}
where $R(z)$ has the series expansion \cite{Cohen92}
\begin{equation} \label{eq:Rgeneral}
	R(z) = VQG_0QV + VQG_0QVQG_0QV +  \cdots .
\end{equation}
$P$ is the projection from $\mathscr H$ to $\mathscr E$ and $Q$ the projector onto its compliment $\bar{\mathscr E}$ and $G_0=(z-H_0)^{-1}$ is the free propagator. The series expansion of $R(z)$ converges provided the energies within $\mathscr E$ are closer to each other than the energy of any state in $\bar{\mathscr E}$ with which they interact. Thus, the non-degenerate basis introduced in \secref{sec:basis} is also necessary here to calculate multi-frequency effects using the resolvent formalism.
The multi-frequency effects are encompassed by $R(z)$ where each term describes an interaction in $\mathscr E$ mediated via virtually excited states in $\bar {\mathscr E}$. These terms are easily evaluated analytically when the two states of interest are close to resonance with each other, but far from resonance with any other state, in which case $z$ is well approximated by their mean energy.

We interpret the effective Hamiltonian in powers of $V$.
The terms of $H_{\mathrm{eff}}$ zeroth order in $V$ are merely the energies of the unperturbed levels of $H_0$. The first order terms in $V$ describe the direct interactions between states of $\mathscr E$. The second order terms, for the interaction discussed in this work, describe interactions which return to the same state they started, via an intermediate state in $\bar{\mathscr E}$. These self interactions shift the levels energy levels of $PH_0P$. Third order terms are between different initial and final states in $\mathscr E$ via two intermediate states in $\bar{\mathscr E}$.

\begin{figure}[!t]
	\begin{picture}(250,215)
		%Vab
		\put(75,202){$V_{ab}=$}
		\put(105,200){\includegraphics[width=40pt]{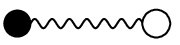}}
		\put(104,190){$\ket{b}$}
		\put(137,190){$\ket{a}$}
		%Raa
		\put(0,128){$R^{(2)}_{aa}\quad=$}
		\put(56,138){$\ket{a}$}
		\put(25,93){\includegraphics[width=40pt]{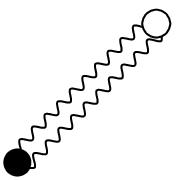}}
		\put(80,128){$+$}
		\put(75,125){\includegraphics[width=40pt]{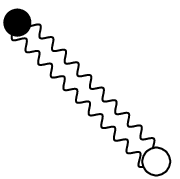}}
		\put(106,115){$\ket{a}$}
		%Rbb
		\put(130,128){$R^{(2)}_{bb}=$}
		\put(160,125){\includegraphics[width=40pt]{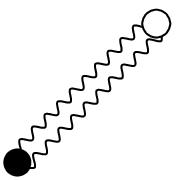}}
		\put(158,115){$\ket{b}$}
		\put(185,128){$+$}
		\put(203,138){$\ket{b}$}
		\put(205,93){\includegraphics[width=40pt]{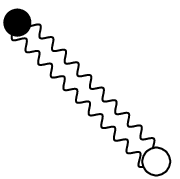}}
		%Rab
		\put(30,43){$R_{ab}^{(3)}\quad=$}
		\put(75,40){\includegraphics[width=40pt]{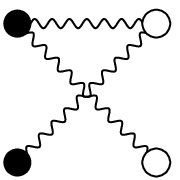}}
		\put(73,30){$\ket{b}$}
		\put(106,30){$\ket{a}$}
		\put(124,42){$+$}
		\put(138,52){$\ket{b}$}
		\put(171,52){$\ket{a}$}
		\put(140,7){\includegraphics[width=40pt]{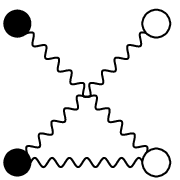}}
	\end{picture}
\caption{\label{fig:q0diagrams} 
Diagrams for one, two and three photon processes for a spin-half in a three frequency field. These diagrams represent the contributions to the effective Hamiltonian describing the resonance between the spin and the middle frequency field mode. The effective Hamiltonian, expanded up to third order interactions, is written $H_{\mathrm{eff}} = PH_0P + PVP + PVQG_0VP + PVQG_0VQG_0VP$. Figure (a) represents $P VP$ between $ \ket{a}=\ket{N, \downarrow}$ and $\ket{b}=\ket{N-j,\uparrow}$. Figure (b) represents the different two photon processes of $R^{2} = PVQG_0VP$ starting and returning to either $\ket{a}$ and $\ket{b}$. Figure (c) represents the three photon processes $R^{3} = PVQG_0VQG_0VP$ between $\ket{a}$ and $\ket{b}$. Note that we have not shown $R^{3}_{ba}$ which is the Hermitian conjugate of $R^{3}_{ab}$ and has the same diagrammatic representation.
}
\end{figure}

To define the effective Hamiltonian at a given resonance we must find the relevant multi-photon processes which modify $PHP$. These processes, described generally by \eqnref{eq:Rgeneral}, are those that begin and end with states in $\mathscr E$ but have all intermediate states in $\bar{\mathscr E}$. Transitions between initial, intermediate and final states can only be those driven by the field modes present. 
Provided the Rabi frequencies are smaller than $\omega_f$, the leading terms are those lowest order in $V$. For three frequencies the processes up to third order are shown diagramatically in \figref{fig:q0diagrams} for the $q=0$ resonance.
The term for each multi-photon process is evaluated from matrix elements of individual transitions involved using \eqnref{eq:approxoffdiagmatrixelements} and the free propagators ($G_0$) of the intermediate states. The later are known from the intermediate state energies evaluated using \eqnref{eq:diagonalmatrixelements}.

We calculate effective Hamiltonians up to third order in $V$, first for the three frequency example considered of \figref{fig:evals3f}, then more generally for a large number of evenly spaced off-resonant frequencies. For convenience we abbreviate the resonant states $\ket{N-(j+q),\tfrac12}$ and $\ket{N, -\tfrac12}$ to $\ket{a}$ and $\ket{b}$ respectively. All energies are expressed relative to the mean energy $E_0=(E_a+E_b)/2=\left [N-(j+q)/2 \right ] \omega_f$ at the resonance under consideration.

We begin by looking at the crossing $q=0$, where the initial and final states are $\ket a =\ket{N-j,\tfrac12}$ and $\ket b=\ket{N,-\tfrac12}$.
We evaluate \eqnref{eq:Heffgeneral} for the processes shown in \figref{fig:q0diagrams}.
Up to first order
\begin{equation} \label{eq:q0firstorderHeff}
	H_{\mathrm{eff}} = P H P = \frac12 \left(
		\begin{array}{cc}
			\Delta_0 & \Omega_j \\
			\Omega_j^* & -\Delta_0
		\end{array}
	\right).
\end{equation}
where $\Omega_j = g_j (\gamma_{N-j}/\gamma_N) \alpha_j\approx g_j \alpha_j$. We have labelled $\Delta$ with the subscript $q=0$ to show which resonance this detuning is from.
At this level of approximation the system behaves like it is coupled only by a single-frequency $j \omega_f$.

The second order term of $R(z)$, $PVQG_0(z)QVP$, shifts the energies of $\ket{a}$ and $\ket{b}$ due virtual excitation of the intermediate states $\ket{N\pm1,-\tfrac12}$ and $\ket{N-j\pm, \tfrac12}$ respectively. 
Thus, at second order the diagonal elements of \eqnref{eq:q0firstorderHeff} are shifted by
\begin{IEEEeqnarray*}{rCl} \label{eq:qzerolevelshifts}
	R^{(2)}_{aa}(z) &=&
		\frac{\abs{\Omega_{j-1}}^2/4}{z+\omega_f+\tfrac12 \Delta_j}
		+\frac{\abs{\Omega_{j+1}}^2/4}{z-\omega_f+\tfrac12 \Delta_j} \IEEEyesnumber \IEEEyessubnumber\\
	R^{(2)}_{bb}(z) &=& 
		\frac{\abs{\Omega_{j-1}}^2/4}{z-\omega_f-\tfrac12 \Delta_j}
		+\frac{\abs{\Omega_{j+1}}^2/4}{z+\omega_f-\tfrac12 \Delta_j}, \IEEEyessubnumber
\end{IEEEeqnarray*}
where $z$ can be approximated by zero.
The resonance is therefore shifted to $\Delta_0=R_{bb}^{(2)}(0)-R_{aa}^{(2)}(0)$.
However, when the amplitudes of the modes $j \pm 1$ are equal the two terms which contribute to each shift cancel to zero when $\Delta_j=0$, so in this case the resonance is unchanged.

The third order contribution modifies the interaction between $\ket{a}$ and $\ket{b}$, coupling them additionally via two intermediate states $\ket{N-j+1,\tfrac12}$ and $\ket{N+1,-\tfrac12}$ or $\ket{N-j-1,\tfrac12}$ and $\ket{N-1,-\tfrac12}$. 
The matrix elements for these terms are
\begin{IEEEeqnarray*}{rCr}
	R^{(3)}_{ab}(z) &=& \frac{\Omega_{j-1}\Omega_j^* \Omega_{j+1}}{8}
		\left[
			\frac{1}{(z-\omega_f-\tfrac12\Delta_j)(z-\omega_f+\tfrac12 \Delta_j)}
		\right.\\
	&&
		\left. +\frac{1}{(z+\omega_f-\tfrac12\Delta_j)(z+\omega_f+\tfrac12 \Delta_j)}
		\right] \IEEEeqnarraynumspace \IEEEyesnumber
\end{IEEEeqnarray*}
and its Hermitian conjugate. At resonance, this third order interaction increases the splitting between $E_a$ and $E_b$ further than the value $\Omega_j$ expected for a single-frequency field.

The same approach can be adopted for the resonances $q=\pm1$ with similar results. To lowest order the matrix elements are $\pm \tfrac12 \Delta_q$, where $\Delta_q=\omega_0-(j+q)\omega_f$. The first order contribution to the interaction matrix elements are $\tfrac12 \Omega_{j+q}$ and its Hermitian conjugate.
The result \eqnref{eq:qzerolevelshifts} for the leading order level shift at $q=0$ can be generalised to
\begin{IEEEeqnarray*}{rCl}
	R^{(2)}_{aa} &=& 
		\sum_{p\neq q, \abs{p}\leq 1} \frac{\abs{\Omega_{j+p}}^2}{4} \frac{1}{z-\tfrac12 \Delta_{j+q}-(q-p)\omega_f},\\
	R^{(2)}_{bb} &=&
		\sum_{p\neq q, \abs{p}\leq 1} \frac{\abs{\Omega_{j+p}}^2}{4} \frac{1}{z+\tfrac12 \Delta_{j+q}+(q-p)\omega_f}.
\end{IEEEeqnarray*}
which not only applies to the cases $q=\pm1$ but for any $q$. Unlike the $q=0$ case, these level shifts do not cancel on resonance. For $q=1$ these terms shift the resonance to negative value of $\Delta_{j+1}$, while for $q=-1$ the shift is to a positive value of $\Delta_{j-1}$.

\begin{figure*}
	\subfloat[Energy Levels \label{fig:q2resonance}]{
		\includegraphics[height=157pt]{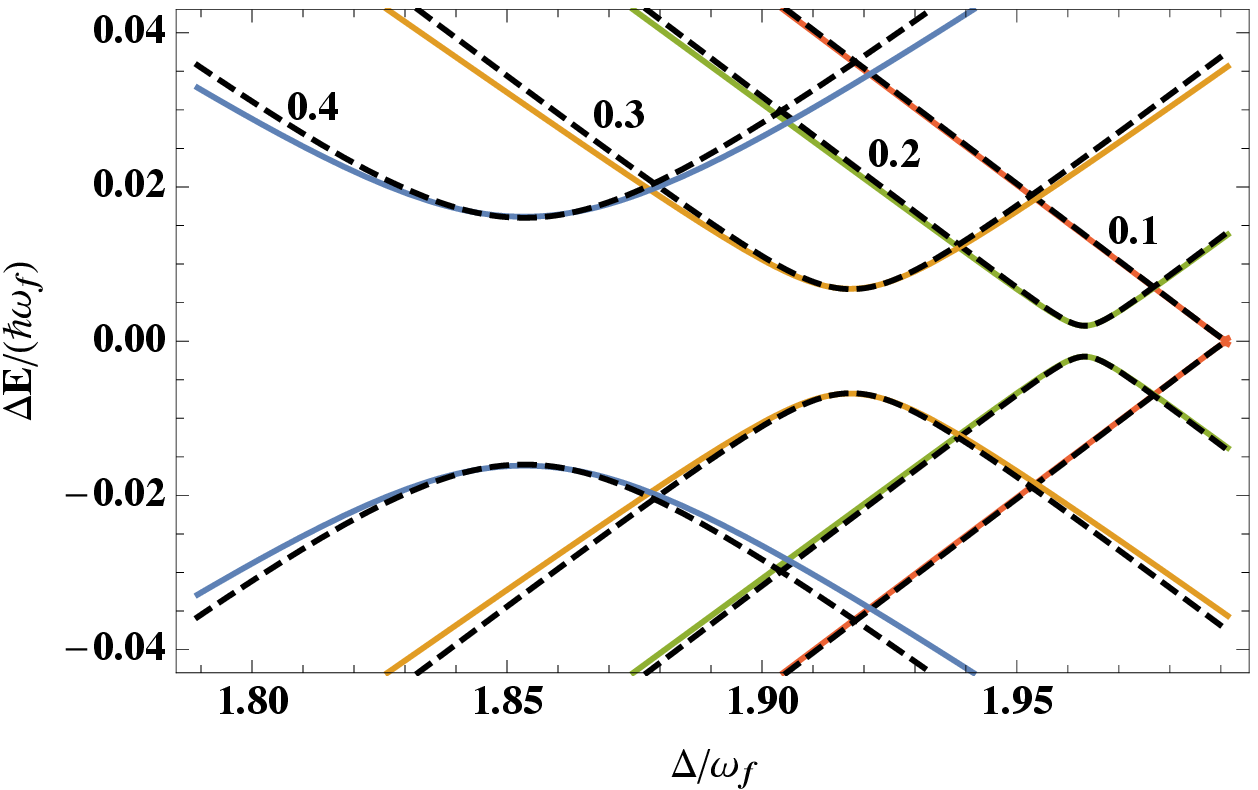}
		}
	\subfloat[Excitation Spectrum \label{fig:q2spectrum}]{
		\includegraphics[height=157pt]{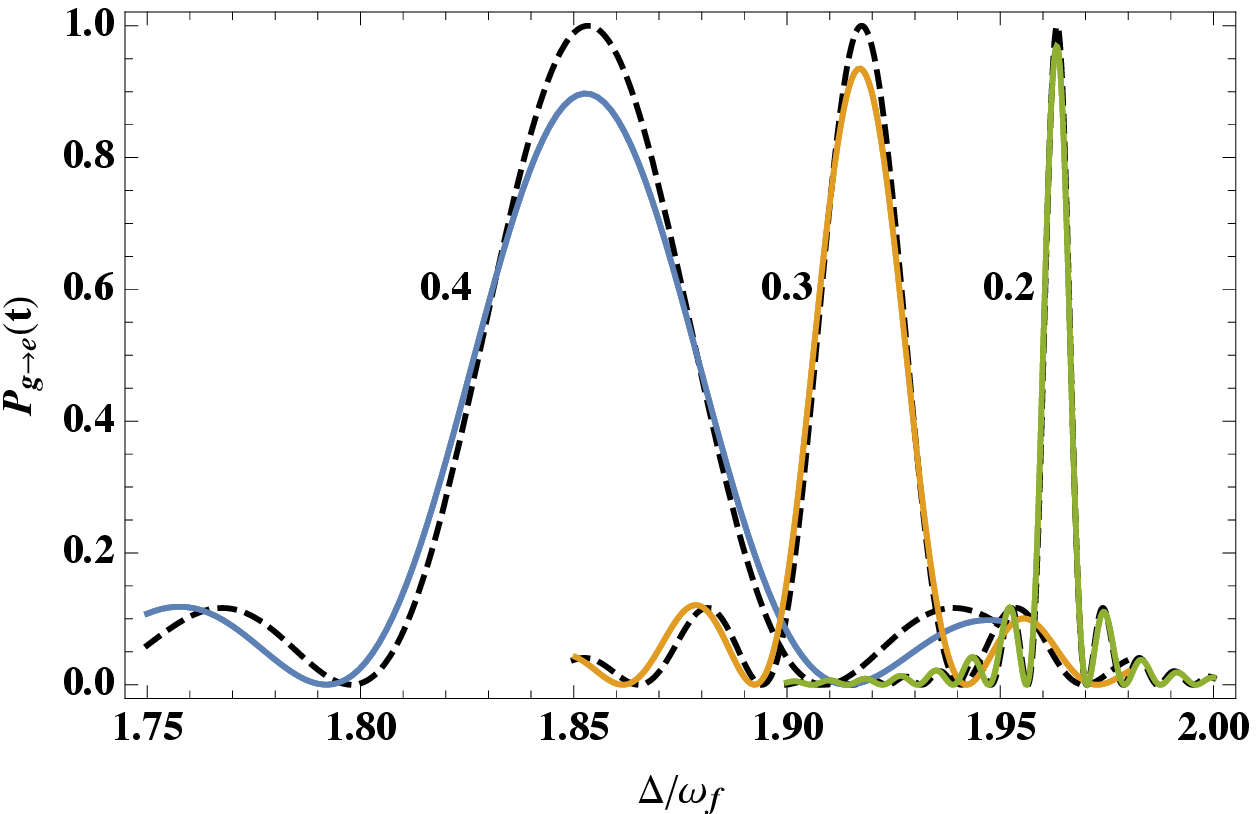}
		}
	\caption{{\label{fig:q2resonance}} 
Dressed state energies (a) and excitation probabilities (b) for a three photon resonance of a spin-half in a three frequency field.
This figure demonstrates the accuracy of the effective Hamiltonian discussed in \secref{sec:effectiveHs}, where dashed lines plot the analytic results of the effective theory, and solid lines the exact numerical results. 
In (a) the three photon avoided crossing becomes larger and shifts left as the Rabi frequencies increase. Here the field frequencies are $j \omega_f$ and $(j\pm1) \omega_f$. The results are plotted against the detuning $\omega_0-j\omega_f$ such that the unshifted resonance occurs at $\Delta = 2$ (i.e $q=2$). All fields have the same Rabi frequencies or equivalently their coherent amplitudes $\alpha_1, \alpha_2$ and $\alpha_3$ are the same. Curves are labelled by these Rabi frequencies in units of $\omega_f$. In (b) this manifests as a broadening and shift left of the peak excitation probability. The excitation probability was calculated for a square pulse of all three frequencies with duration $\tilde \Omega_{\mathrm{eff}} t = \pi$ where $\tilde \Omega_{\mathrm{eff}}=2 R_{ab}^{(3)}$ is the effective three-photon Rabi frequency given in \eqnref{eq:q2omegaeff}.
}
\end{figure*}

For level crossings where $\abs{q}>1$ there is no direct coupling between the states $\ket{a}$ and $\ket{b}$ within the resonant subspace $\mathscr E$. For $q=\pm2$ or $q=\pm3$ the resonant states are only coupled by third and higher order processes. In general we find that at the $q^{\mathrm{th}}$ resonance, between states $\ket{a}=\ket{N-(j+q),\tfrac12}$ and $\ket{b}=\ket{N,-\tfrac12}$, the lowest order interaction is of order $q$ in $V$ when $q$ is odd, and of order $q+1$ when $q$ is even. For odd $q$ the effective interaction matrix element between the resonant states is
\begin{IEEEeqnarray*}{rCl}
	R_{ab}^{(q)} &= 
		\frac{\left({\Omega_{j+1}^{\frac12(q+1)}}\right)^* \Omega_{j-1}^{\frac12 (q-1)}}{2^q} 
		&\prod_{n=1}^{\tfrac12(q-1)} \left(\frac{1}{z+\tfrac12 \Delta_{j+q}+2 n \omega_f} \right.\\
	&& \left.
		\times \frac{1}{z-\tfrac12 \Delta_{j+1} - 2 n \omega_f} \right). \IEEEyesnumber
\end{IEEEeqnarray*}
On resonance ($\Delta_q=0$), and approximating $z$ with zero, this effective interaction matrix element simplifies to
\begin{equation}
	R_{ab}^{(q)} \approx 
		\left(\frac{\Omega_{j+1}^*}{2}\right)^{\frac12(q+1)}  \left(\frac{-\Omega_{j-1}}{2} \right)^{\frac12 (q-1)}
		\frac{1}{(2 \omega_f)^{q-1} \left(\frac{q-1}{2}!\right)^2}
\end{equation}
This shows the multi-photon resonant interaction strength scales generally as
\begin{equation}
	R_{ab}^{(q)} \sim 8 \Omega \left(\frac{\Omega}{2 \omega_f}\right)^q \left(\frac{q}{q!}\right)^2
\end{equation}
with a characteristic Rabi frequency $\Omega\sim \Omega_j \forall j$.

For even $q$ there are $q+1$ different processes which contribute to the interaction matrix element, each of order $q+1$ in $V$. While the interaction processes are more numerous for even $q$ resonances, the coupling is typically weaker than that for $q-1$ unless the number of processes exceeds the relevant ratio $\Omega / \Delta$. Due to the multiple processes involved, the lowest order multi-photon matrix element for even $q$ is given by the more complicated expression
\begin{widetext}
\begin{IEEEeqnarray*}{rCl}
	R_{ab}^{(q+1)} =
		&\sum_{s=1}^{q/2} &\left \{
			 \frac{\Omega_j^*}{2} \frac{\Omega_{j+1}/2}{z-(q-2s+1)\omega_f-\tfrac12 \Delta_{j+q}} 
			\left[ \prod_{n=1}^{s-1} \frac{\Omega_{j-1}^*/2}{z+2 n \omega_f+\tfrac12 \Delta_{j+q}}
				\frac{\Omega_{j+q}/2}{z-q \omega_f + (2n+1)\omega_f-\tfrac12 \Delta_{j+q}} \right] \right. \\
	&& \left. \times
			\frac{\Omega_{j+1}/2}{z+(2s-1)\omega_f+\tfrac12 \Delta_{j+q}} 
			\left[
				\prod_{n=1}^{q/2-s-1} \frac{\Omega_{j-1}^*/2}{z+q \omega_f - (2n-1)\omega_f+\tfrac12 \Delta_{j+q}}
				\frac{\Omega_{j+1}/2}{z-2 n\omega_f-\tfrac12 \Delta_{j+q}}
			\right] \right\} \\
	&+ \sum_{s=0}^{q/2} & \left\{
		\frac{\Omega_j}{2}
		\left[
			\prod_{n=1}^{s} \frac{\Omega_{j+1}/2}{z+2 n \omega_f+\tfrac12 \Delta_{j+q}}
			\frac{\Omega_{j-1}^*/2}{z-q \omega_f + (2n-1)\omega_f-\tfrac12 \Delta_{j+q}}
		\right] \right. \\
	&&\left. \times
		\left[ \prod_{n=s}^{q/2 -1} \frac{\Omega_{j+1}/2}{z+(2n+1)\omega_f+\tfrac12 \Delta_{j+q}}
			\frac{\Omega_{j-1}^*/2}{z-q \omega_f +2 n \omega_f - \tfrac12 \Delta_{j+q}}
		\right]
		\right\}.
\end{IEEEeqnarray*}
\end{widetext}
This expression is greatly simplified when applied to an individual resonance. For example the third order interaction for $q=2$ has
\begin{IEEEeqnarray*}{rCl}
	R_{ab}^{(3)} &=& \frac{\Omega_j}{2} \frac{\Omega_{j-1}^*}{2} \frac{\Omega_{j+1}}{2} 
		\left[ \frac{-1}{(\Delta_2+2\omega_f)(\Delta_2+\omega_f)}\right. \IEEEyesnumber \label{eq:q2omegaeff} \\
	&& \left.
		+  \frac{-1}{(\Delta_2+\omega_f)(\Delta_2+2 \omega_f)} +  \frac{-1}{(\Delta_2+\omega_f)(\Delta_2+ \omega_f)} \right],
\end{IEEEeqnarray*}
with $z\approx0$.
On resonance this simplifies to $\Omega_j \Omega_{j-1}^* \Omega_{j+1}/4$.

The effective Hamiltonian for each two dimensional subspace is easily solved to find the dressed states and their energies, which in general are
\begin{equation} \label{eq:Heffdressedenergies}
	E_{\pm} = \frac12 \sqrt{\left[\Delta_q-(R_{bb}-R_{aa})\right]^2+(2 R_{ab})^2}.
\end{equation}
Thus the effective Rabi splitting at the multiphoton resonances is $2 R_{ab}$. 
An atom, initially in its ground state therefore undergo Rabi oscillations
\begin{equation} \label{eq:rabiosc}
	\left \langle \tfrac12 \vert  \psi (t) \right \rangle = 
		\frac{\Omega_{\mathrm{eff}}}{\widetilde \Omega_{\mathrm{eff}}} \sin \left(\widetilde \Omega_{\mathrm{eff}} t/2 \right),
\end{equation}
where $\Omega_{\mathrm{eff}}=2R_{ab}$ and $\widetilde \Omega_{\mathrm{eff}} = \sqrt{\left[\Delta_q-(R_{bb}-R_{aa})\right]^2+\Omega_{\mathrm{eff}}^2}$.
Figure \ref{fig:q2resonance} shows the dressed energies calculated in this way agree well with the numerical approach given in \secref{sec:calculations}.
Figure \ref{fig:q2spectrum} shows the Rabi oscillations of this two-level subsystem calculated from the square modulus of \eqnref{eq:rabiosc}, and for comparison, the same result calculated numerically. As $\Omega_k$ decrease the analytic approximation becomes more accurate. For larger $\Omega_k$ there is a leakage of the population to states outside the subspace considered, which is accounted for in the numerical results. In other work, we find simple analytic expression which accurately describe the time evolution in this case \cite{yuen2018analytic}.

\section{Weak fields and the boundary between classical and quantum electromagnetic fields} \label{sec:quantumclassicalboundary}

\begin{figure*}
	\begin{tabular}{lr}
		\subfloat[Energy Levels \label{fig:alphaplot1}]{
			\includegraphics[width=240pt]{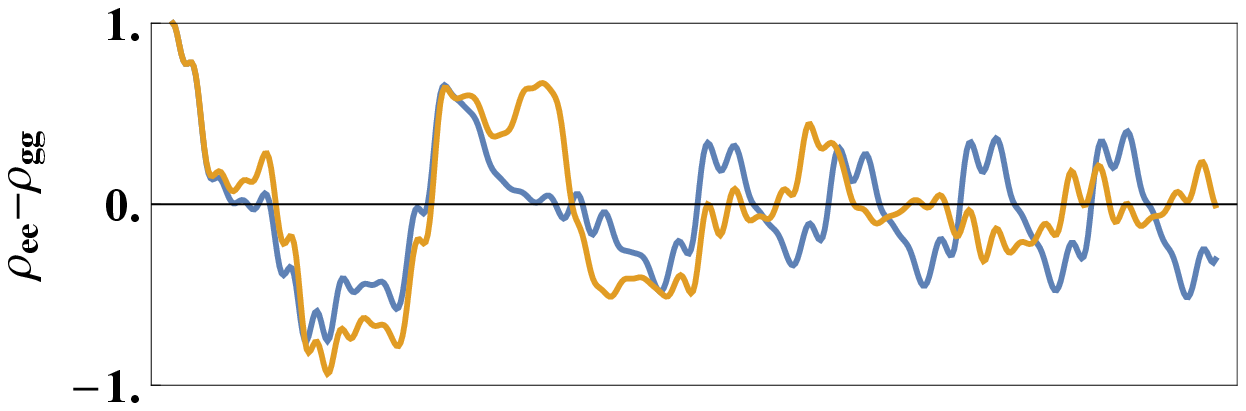}
			\put(-70,60){$\alpha = 2, \ \omega_0=2$}
		} 
		& 
		\subfloat[Energy Levels \label{fig:g0plot1}]{
		\includegraphics[width=240pt]{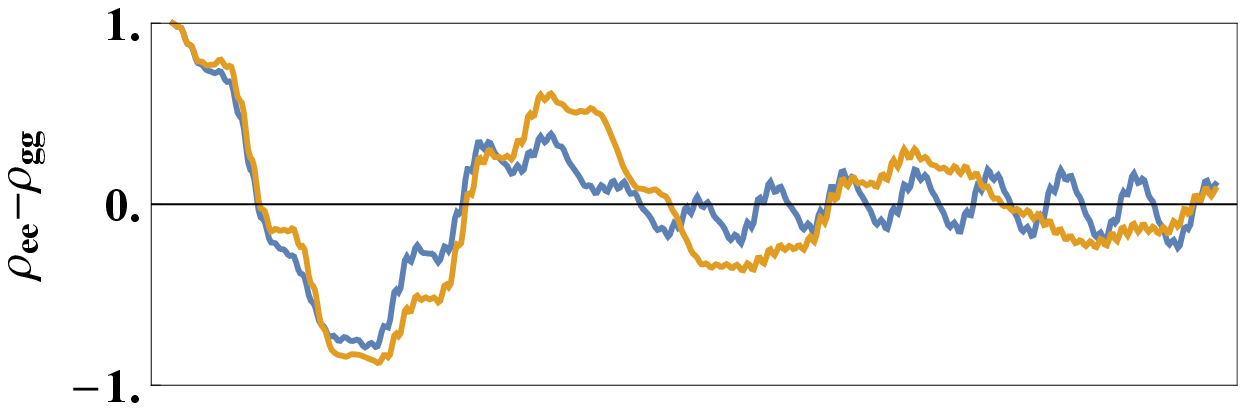}
		\put(-90,60){$\omega_0=2, \ g_0/\bar \omega = 10^{-2}$}
		} 
		\\
		\subfloat[Energy Levels \label{fig:alphaplot2}]{
			\includegraphics[width=240pt]{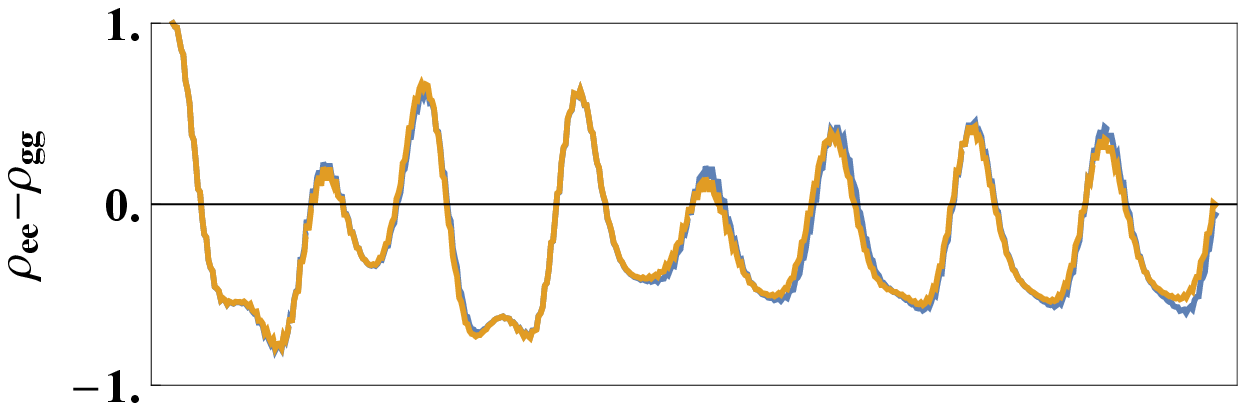}
			\put(-70,60){$\alpha = 1, \ \omega_0=11$}
		} 
		& 
		\subfloat[Energy Levels \label{fig:g0plot2}]{
			\includegraphics[width=240pt]{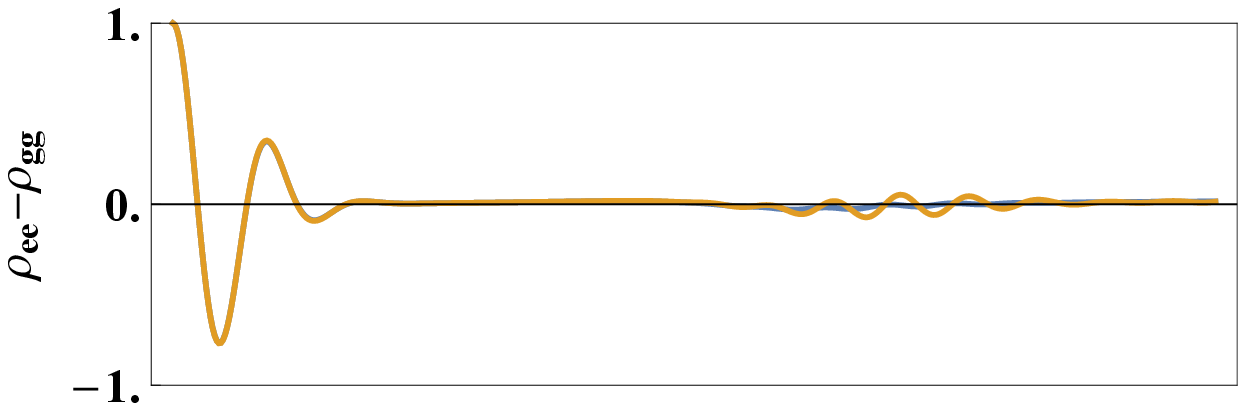}
			\put(-90,60){$\omega_0=2, \ g_0/\bar \omega = 10^{0.5}$}
		} 
		\\
				\subfloat[Energy Levels \label{fig:alphaplot3}]{
			\includegraphics[width=240pt]{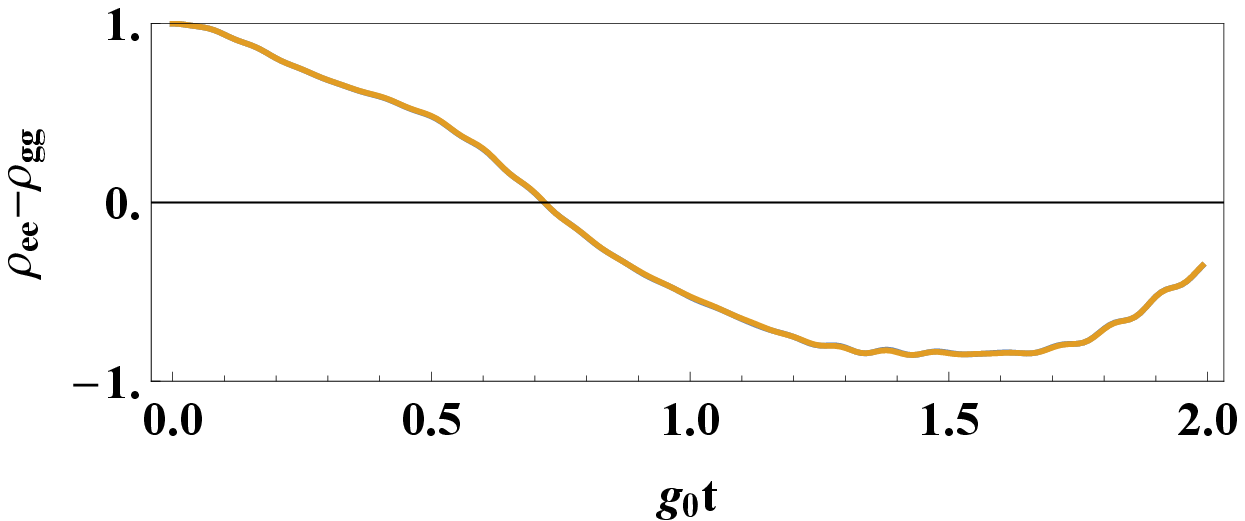}
			\put(-70,85){$\alpha = \frac{1}{\sqrt{2}}, \ \omega_0=3$}
		} 
		& 
		\subfloat[Energy Levels \label{fig:g0plot3}]{
			\includegraphics[width=240pt]{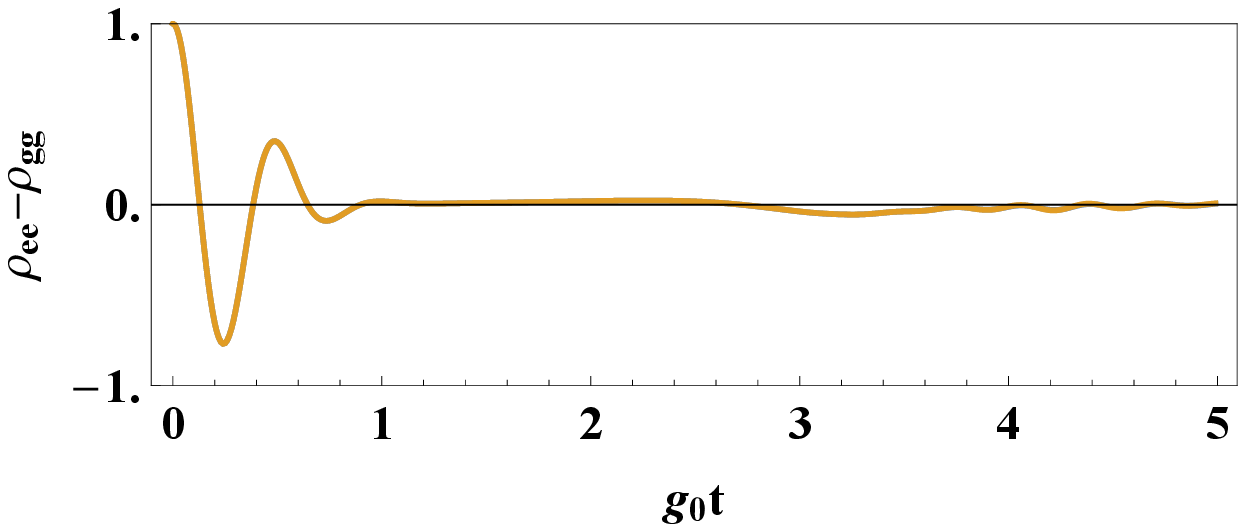}
			\put(-90,85){$\omega_0=9, \ g_0/\bar \omega = 10^{0.5}$}
		} 
	\end{tabular}
	\caption{\label{fig:popinversioncollapse} 
Multi-frequency collapse and revivals of a spin-half system driven by two frequencies. Each plot shows the spin-half population inversion calculated in the non-degenerate basis  (orange lines) and Fock basis (blue lines) for comparison. The multi-frequency dynamics are clearly more complex than for a single-frequency coherent driving field. Collapse and revivals are predicted by the non-degenerate formalism because it accounts for the state-dependent action of the field operator, unlike semi-classical approaches such as Floquet theory. Results are shown for a variety of different coherent amplitudes $\alpha_1=\alpha_2=\alpha$, coupling strengths $g_0/\bar \omega$ (normalised to the mean field freqeuncy), and mean field frequencies. The different plots show that as one moves away from the parameter regime $\alpha \sim 1, \ g_0/\bar \omega \sim 1$ or towards higher $\bar \omega_0$ that the accuracy of results calculated in the non-degenerate basis increases. Calculations in this basis are significantly more efficient than in the Fock basis as the system size does not grow exponentially with the number of field modes. A full comparison between the non-degenerate basis and Fock basis (exact results) is given in \figref{fig:l2comparisons}.
Energy is in units of $\omega_f$ which is also the spacing between the two frequency modes, and time is shown in units of $1/g_0$.
}
\end{figure*}

\begin{figure*}
	\subfloat[\label{fig:alphavsf0plot}]{
		\includegraphics[width=250pt]{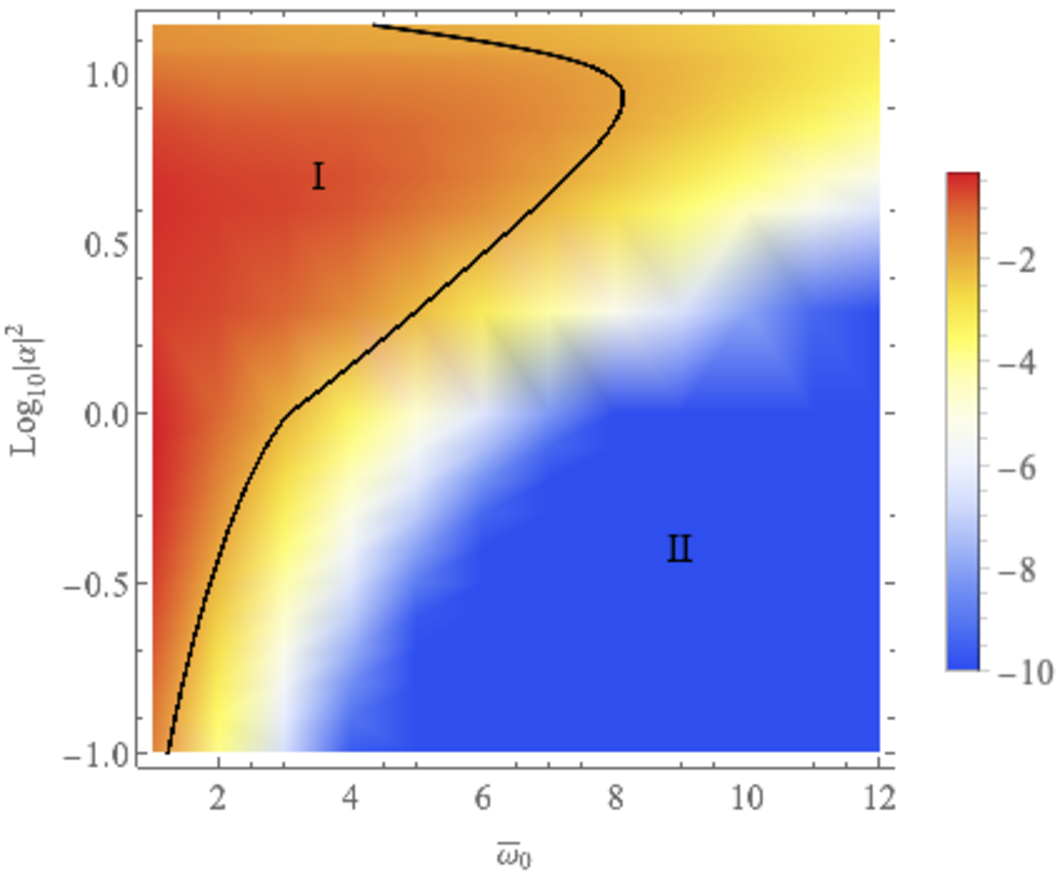}
		}
	\subfloat[ \label{fig:g0vsf0plt}]{
		\includegraphics[width=250pt]{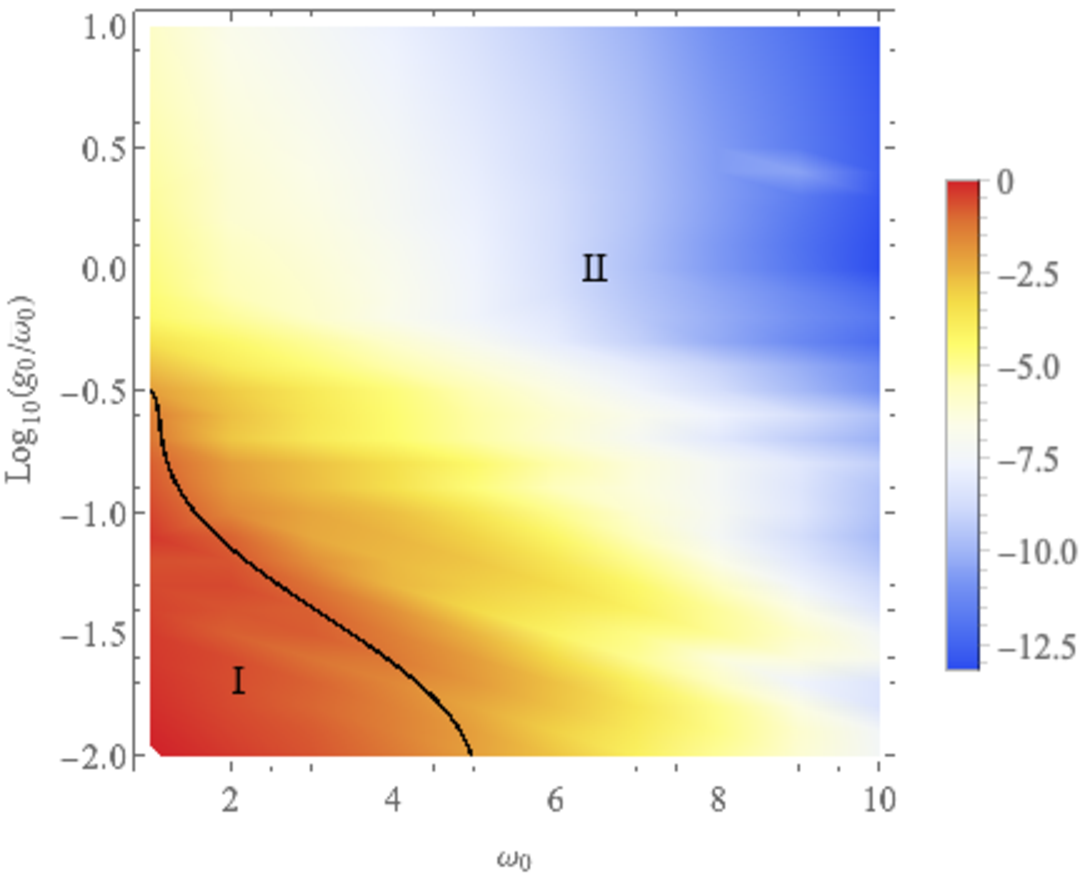}
		}
	\caption{{\label{fig:l2comparisons}} 
%
%\begin{figure*}
%	\subfloat[Alpha versus  \label{fig:alphavsf0plot}]{
%		\includegraphics[width=250pt]{alphavsf0plot}
%		}
%	\subfloat[Interaction strength versus \label{fig:g0vsf0plt}]{
%		\includegraphics[width=250pt]{g0vsf0plt}
%		}
%	\caption{{\label{fig:l2comparisons}} 
This figure shows there are two regimes of multi-frequency quantum dynamics for a spin-half driven by a two frequency field.
These regimes occur in the parameter space of the mean photon number in each mode ($\abs{\alpha}^2$), mean field frequency $\bar \omega_0$, and interaction strength ($g_0/\bar \omega$) scaled by this frequency.
These regimes are highlighted by the discrepancy (indicated by colour) between the population inversion calculated in the non-degenerate and Fock bases. In region II, the non-degenerate basis produces accurate results as the state of the field remains separable into different modes. In region I the discrepancy occurs because the field modes become entangled, requiring a full description in the Fock basis. The discrepancy was quantified using the square difference between the population inversion results up to time $g_0 t = 1$, as given by \eqnref{eq:shrodrabi}. The colour indicates this discrepancy on log scale with base 10. The boundary between the regions was drawn at a 1\% discrepancy. In (a) $g_0/\bar \omega $ is constant at 0.1, and in (b) $\abs{\alpha}^2 = 5$. Examples of the population inversion at six different points in the parameter space shown here are given in \figref{fig:popinversioncollapse}.
}
\end{figure*}

The electromagnetic fields considered here are in a product of coherent-states, which are the quantum states closest to coherent classical fields. We know however, that they differ from classical fields, and this differenced is most pronounced for weak fields where the mean photon number in each mode is small $\abs{\alpha_k} \sim 1$. For a single-frequency interacting with a spin-half system, a collapse of the Rabi oscillations is observed due dependence on the interaction strength on the square root of the photon number. Some time later revivals occur, but never complete since the frequency with which different pairs of atom-photon states oscillate is irrationally related \cite{Eberly1980}.
Thus, collapse and revivals are one hallmark of a quantised coherent field.
We investigate the limits in which our non-degenerate theory accurately encapsulates the collapse of Rabi oscillations. Remarkably, we find that the our theory accurately predicts the collapse in the deep strong coupling regime in addition to the limit where the coherent fields are highly excited.

When multiple frequencies are present additional mechanisms effect the process of collapse. Additional frequencies change how the different components of oscillating spin-field system dephase. When additional frequencies are far from resonance compares to the interaction strength they drive virtual excitations which shift and the resonantly interacting levels providing further dephasing. 
When the interaction strength becomes comparable to the detuning between the field modes, the spin system is driven by many modes without any single mode dominating its dynamics. 
In turn the different modes of the field become entangled via the spin.
As the interaction strength is increase beyond the mode spacing the multiple modes start to drive the spin collectively, each acting in a similar way.

To investigate the multi-mode dynamics and validate our theoretical formalism we calculate the time evolution for the multi-mode Rabi model numerically, both in the Fock basis and our non-degenerate basis, and compare the time evolution of the spin inversion $\rho_{ee}-\rho_{gg}$. The multimode Rabi model considered is
\begin{IEEEeqnarray}{C} \label{eq:shrodrabi}
	i \partial_t \ket{\psi} = \left[\tfrac12 \omega_0 \sigma_z + \sum_i a_i^{\dagger} a_i \omega_i + g_i (a_i^{\dagger}+a_i) \sigma_x \right] \ket{\psi} \IEEEeqnarraynumspace
\end{IEEEeqnarray}
with coupling strength $g_i = \sqrt{\omega_i/\omega_0} g_0$, and $\omega_0$ is the lowest field frequency which we set resonant spin. Thus the coupling strength grows with $\sqrt{\omega}$ as is the case for the electromagnetic field, and is parameterised by $g_0$. We calculate $\ket{\psi(t)}$ for fields with two three frequencies with mode space $\Delta=1$ in dimensionless units. The initial state is a product of coherent-states, each with equal amplitude $\alpha$, with spin up. We compare $\rho_{ee}-\rho_{gg}$ calculated in the Fock and non-degenerate bases for a range of $g_0, \omega_0 $ and $\alpha$. We quantify the comparison between results in each basis using the $L_2$ measure
\begin{IEEEeqnarray}{rCl} \label{eq:l2measure}
	L_2[f(t),g(t)] = g_0 \int_0^{1/g_0}  \abs{f(t)-g(t)}^2 \mathrm d t.
\end{IEEEeqnarray}
This integral is over the interval $t=0$ to $t=1/g$ - the characteristic time of the collapse for single-frequency Rabi oscillations \cite{puri1986}.

Figure \ref{fig:l2comparisons} compares the spin population inversion calculated in the Fock and non-degenerate basis over a range of field amplitudes $\alpha$, coupling strengths $g_0$, and field frequencies $\bar \omega_0 = \omega_0 + \Delta/2$. Figure \ref{fig:popinversioncollapse} shows typical population inversions for different $\alpha$ and $\omega_0$ with $g_0=\bar \omega/10$ in figures (a) to (c)), and different $g_0=\bar \omega$ and $\omega_0$ with $\abs{\alpha}=5$ in figures (d) to (f).
In all cases there there is close agreement between Fock and non-degenerate basis calculations for times $g_0 t \ll 1$. When $\bar \omega \sim 1$, $\alpha \sim 1$ and $g0/\bar \omega_0 \sim 10^{-1}$ (ultra strong coupling), we find the results are quite different already by $g_0 t \sim 1$. Typical examples are shown in \figref{fig:alphaplot1} and \ref{fig:g0plot1}. We refer to this region of parameter space as region I, and the region outside this as region II. The level of agreement is quantified by their $L_2$ measure which is plotted in \figref{fig:l2comparisons} and the boundary between regions I and II is indicated by the contour $L_2 = 10^{-3}$. For parameters in region II, the agreement becomes remarkably good even for longer times. Figures \ref{fig:alphaplot2}, \ref{fig:g0plot2}, \ref{fig:alphaplot3} and \ref{fig:g0plot3} show the population inversion typical of region II.

The strongest discrepancy between the Fock and non-degenerate basis calculations is for small $\omega_0$ where it is comparable to the frequency difference $\Delta$, small $g_0/\Delta$, and mean photon numbers $\abs{\alpha} \sim 1$. This arises because the non-degenerate basis does not cover Fock space since it represents many non-degenerate Fock states by a single state. The action of the field operators on this state is derived from the action on each underlying Fock state together with the set of amplitudes with which each contributes to the non-degenerate state. The action is correct provided that the field remains in a coherent-state. Gradually, the Fock states which contribute to a non-degenerate state will dephase. The rate of dephasing will be most significant when both field modes interact strongly with the spin but the two modes remain distinct such that their dynamics are different.

When the dephasing between components of the non-degenerate states is slow the non-degenerate basis yields accurate results.
This is the case when $g_0$ becomes much larger than the mode spacing such that the difference in detuning between the spin and field modes becomes irrelevent. The dephasing is also slows as $\bar \omega_0$ increase for fixed $\Delta$. A perturbative expansion of the interaction shows that there are fewer low order processes which resonantly couple multi-mode Fock states when their frequencies approach.

In the limit of large photon number $\bar n \gg 1$ it is common to ignore the $n$-dependence of the interaction between different Fock states since its variance $\mathrm {var}(n)=\sqrt{\bar n}\ll \bar n$. While this is widely accepted, its universal application is limited to short times much shorted than $1/g_0$ - the timescale of Rabi oscillation collapse and is caused by the $n$-dependence of interactions. For multiple frequencies we give a similar approximation where the ratios $\gamma_N/\gamma_{N\pm k}$ are ignored, the variance of $N$ for this approximation being $\sigma_N$ and not $\sqrt{N}$. Figure \ref{fig:alphavsf0plot} suggests that on a timescale $1/g_0$ the non-degenerate basis is remains suitable for accurate estimation of the spin population in the limit of large $\bar N$, but that this limit is approached slowly, compared with the increase in accuracy of the non-degenerate basis calculations for increasing frequency or coupling strength.

The comparison between the spin-dynamics calculated in the Fock and non-degenerate bases highlight two distinct regimes of multi-frequency light-matter interactions with coherent fields. Regime I: For weak fields and small $g_0/\Delta$ the dephasing between degenerate Fock states becomes important and leads to entanglement between the field modes via the spin. Regime II: For strong fields, or for weak fields and large $g_0/\Delta$, each mode remains in a coherent-state, and although the field and spin become entangled, the entanglement between modes is negligible. In the later regime, the dynamics of the field can be accurately represented in the non-degenerate basis. The semi-classical limit applies only for short times compared to $1/g_0$. The quantum dynamics then emerge for longer times, in one of these two regimes. In regime I the entanglement between the modes require a full quantum picture. Regime II lies between the quantum and classical limits, whereby its dynamics depend on the level of excitation of the field, but the field remains factorizable.

\section{Conclusion} \label{sec:conclusion}

We have developed a formalism which tackles the complexity of multi-frequency atom-photon interactions when each field mode is well described by a coherent-state.
We demonstrate its utility for strong strong fields and remarkably, for a wide range of parameters for weak fields where the quantum mechanical nature of the dynamics are important.

This formalism utilises a subspace of Fock space, spanned by a non-degenerate basis. This basis addresses the issue of level degeneracy encountered when working in the Fock basis. We have derived the effect of the photon creation and annihilation operators on these basis states including the asymmetry between them associated with quantum fluctuations. 
For highly excited coherent-states, where the photon number in each mode is large compared to unity, we show these fluctuations are small and can be ignored.
We have demonstrated that by working in this basis, accurate calculations of the multi-frequency dressed states can be performed. Applying the resolvent formalism in this basis we have derived effective Hamiltonians which includes analytic expression for the level shifts and multi-photon interactions for an atom in a multi-frequency field.

For weak fields, where quantum effects are significant, we have compared the long time dynamics predicted by our formalism to exact calculations in the Fock basis. By investigating the parameter range for which our formalism is applicable we have identified two different quantum regimes for the collapse and revival of Rabi oscillations of a spin-half system in a multi-frequency field. One of these covers a large region in parameter space where some quantum aspects of the field are essential for the collapse, but the fields do not become entangled. Our formalism is uniquely capable of addressing problems in this regime.

\begin{acknowledgments}
This work was supported through the UK National Quantum Technologies Programme (NQIT hub, EP/M013243/1) and the EU H2020 Collaborative project QuProCS (Grant Agreement No. 641277). 
%B.Y. and A.K. express gratitude to Ezra Kassa for helpful comments on this manuscript. B.Y. would like to additionally thank Candadi Sukumar, Kathrin Luksch and Rian Hughes for helpful discussions during the preparation of this manuscript.
The authors would like to thank Axel Kuhn, Candadi Sukumar, Kathrin Luksch and Rian Hughes for helpful discussions during the preparation of this work, and Ezra Kassa for comments on the manuscript.
\end{acknowledgments}

\bibliography{MultiFrequencyAtomPhotonInteractions}
%\bibliography{bibliography}

\appendix

\section{Partition of Fock space} \label{ap:partition}

We show that the Fock space for a field with rationally related frequency modes $\mathscr H_{F}$ is partitioned by the subspaces $\mathscr{E_N}$.
Each subspace $\mathscr E_N$ is spanned by the set of all product states $\ket{ \{ N ; n_k \} } = \ket{n_1,n_2,...}$, where each product state $\ket{ \{ N; n_k \} }$ is defined by a set of integers for which $\sum_{k=1}^{\infty} k n_k = N$.
\begin{proof}
Given any set of integers $\{ N; n_k \}$ such that $\sum_k k n_k = N$, 
\begin{IEEEeqnarray*}{rCl}
	H_F \ket{\{ N; n_k\}} 
		&=& \sum_j j \hbar \omega_f a_j^{\dagger} a_j {\prod_k}^{\otimes} \ket{n_k} \\
		&=& \sum_j n_j \hbar \omega_f \ket{\{ N; n_k \}} \\
		&=& N \hbar \omega_f \ket{\{ N; n_k \}}
\end{IEEEeqnarray*}
Hence the set of all such product states, $\left \{ \ket{\{ N; n_k \} } \right \} \subset \mathscr E_N$.

Conversely, given any state $\ket{\psi} \in \mathscr E_N$, it can be expanded on the basis of product states $\ket{ \{ n_k \} } = {\prod_k}^{\otimes} \ket{n_k}$ which span $\mathscr H_F$;
\begin{equation*}
	\psi = \sum_{\{ n_k \}} c_{ \{ n_k \} } {\prod_k}^{\otimes} \ket{n_k}.
\end{equation*}
Since $\ket{\psi} \in \mathscr E_N$,
\begin{equation} \label{eq:HFpsi1}
	H_F \ket{\psi} = N \hbar \omega_f \sum_{\{ n_k \}} c_{ \{ n_k \} } {\prod_k}^{\otimes} \ket{n_k},
\end{equation}
but
\begin{IEEEeqnarray*}{rCl}
	H_F \ket{\psi} 
		&=& \sum_j j \hbar \omega_f a_j^{\dagger} a_j \sum_{\{ n_k \}} c_{ \{ n_k \} } {\prod_k}^{\otimes} \ket{n_k} ,\\
		&=& \hbar \omega_f \sum_{\{n_k \}} c_{\{n_k \}} \sum_ k k n_k {\prod_k}^{\otimes} \ket{n_k},\IEEEyesnumber \label{eq:HFpsi2}
\end{IEEEeqnarray*}
where the $n_k$ under the sum over $k$ correspond to each $n_k \in \{ n_k \}$.
Subtracting \eqnref{eq:HFpsi2} from \eqnref{eq:HFpsi1} and cancelling the factor $\hbar \omega_f$,
\begin{equation}
	\sum_{\{n_k \}} c_{\{n_k \}} \left( N-\sum_k k n_k	\right) \ket{\{ n_k \}} = 0.
\end{equation}
Since the basis states $\ket{\{n_k \}}$ are linearly independent, either $c_{\{n_k\}} = 0$ or $N - \sum_k k n_k = 0$ for each configuration $\{ n_k \}$. However, the state $\ket{\psi} \neq 0$, so there is a least one coefficient for which $c_{\{n_k \}} \neq 0$. For each non zero coefficient $c_{\{ n_k \}}$, we must have
\begin{equation}
	\sum_k k n_k = N.
\end{equation}
Hence for each basis state $\ket{\{ n_k \} }$ in the expansion of $\ket{\psi}$, we must have $\ket{\{ n_k \}} = \ket{ \{N;n_k \}} \in \mathscr E_N$. Therefore $\mathscr E_N \subset \left \{ \ket{\{ N; n_k \} } \right \}$. Thus $\mathscr E_N = \left \{ \ket{\{ N; n_k \} } \right \}$.
\end{proof}

To prove the subspaces $\mathscr E_N$ partition $\mathscr H_F$ we note that $\sum_k k n_k$ is always a positive integer since $k$ is an integer greater than zero and $n_k$ is any positive integer. Hence given any state $\ket{\{ n_k\}}, \ \exists \ \mathscr E_N$ such that $\ket{\{ n_k \}} \in \mathscr E_N$. Conversely, given any $N$, $\exists$ a state $\ket{\{n_k \}}$ such that $\ket{\{n_k \}} \in \mathscr E_N$, i.e. the state for which $n_k=\delta_{k,N}$ where $\delta_{k,N}$ is the Kronecker delta symbol.

\section{$\gamma_N^2$ distribution} \label{ap:gammaapprox}

We now show the distribution $\gamma_N^2$ tends towards a Gaussian distribution when the mean excitation of each field mode $\abs{\alpha_k}^2$ is large. Let the standardised variables $Z=(N-\bar N) / \sigma_N$ and $z_k = k(n_k-\abs{\alpha_k}^2)/\sigma_N$. Furthermore, let the constants $\lambda_k$ be  defined such that $\abs{\alpha_k}^2 = \lambda_k \bar N$.
The characteristic function for $Z$ is given by $\chi_Z(t) = \langle e^{i t Z} \rangle$. Expanding the expectation value,
\begin{IEEEeqnarray*}{rCl}
	\chi_Z(t) 
		&=& \left \langle \exp \left[ i \frac{t}{\sigma_N} (N-\bar N)/\sigma_N \right] \right \rangle \\
		&=& \bra{\{ \alpha_k \}} \exp \left( i \frac{t}{\sigma_N} 
			\sum_k k \hat n_k - k \abs{\alpha_k}^2 \right) \ket{ \{ \alpha_k \}} \\
		&=& \prod_k \bra{ \alpha_k}  \exp \left[ i \frac{k t}{\sigma_N} 
			(\hat n_k - \abs{\alpha_k}^2) \right] \ket{ \alpha_k} \\
		&=& \prod_k \chi_{z_k}(t), \IEEEyesnumber \label{eq:chiZexpansion}
\end{IEEEeqnarray*}
where $\chi_{z_k}(t) = \langle \exp(i t z_k) \rangle$. These can be expanded as
\begin{IEEEeqnarray*}{rCl}
	\chi_{z_k}(t)
		&=&\bra{\alpha_k} \exp \left [ i \frac{k t}{\sigma_N} \left( \hat n_k - \abs{\alpha_k}^2 \right ) \right ] \ket{\alpha_k} \\
		&=& e^{-\abs{\alpha_k}^2} e^{-i \frac{k t}{\sigma_N} \abs{\alpha_k}^2} \\
		&& \quad \times \sum_{n_k,m_k}  \frac{(\alpha_k^*)^{m_k} \alpha_k^{n_k}}{\sqrt{m_k! n_k!}}
			\bra{m_k} e^{i \frac{k t}{\sigma_N} \hat n_k} \ket{n_k} \\
		&=& e^{-\abs{\alpha_k}^2} e^{-i \frac{k t}{\sigma_N} \abs{\alpha_k}^2} \sum_{n_k} 
			\frac{\left( \abs{\alpha_k}^2 e^{i \frac{k t}{\sigma_N}} \right)^{n_k}}{n_k!} \\
		&=& \exp \left[ \abs{\alpha_k}^2 \left(  e^{i \frac{k t}{\sigma_N}}-1-i \frac{k t}{\sigma_N} \right) \right ].
		\IEEEyesnumber \label{eq:chizkexpansion}
\end{IEEEeqnarray*}
We show that in the limit $\{\abs{\alpha_k}^2 \}, \bar N \rightarrow \infty$, the characteristic functions $\chi_{z_k}(t)$ converge pointwise to Gaussian distributions. Let us redefine the distribution means and variances as $\abs{\alpha_k}^2 = \lambda_k \bar N$, where the $\lambda_k$ are held constant in the limit $\{ \abs{\alpha_k}^2 \}, \bar N \to \infty$, such that the ratios $\abs{\alpha_j}/\abs{\alpha_k}$ remain fixed. Furthermore, we define the finite constant $\sigma^2 = \sum_k k^2 \lambda_k$, such that $\sigma_N^2 = \sigma^2 \bar N$. In terms of $\lambda_k, \sigma$, and $\bar N$, $\chi_{z_k}(t)$ becomes
\begin{IEEEeqnarray}{rCl}
	\chi_{z_k}(t) &=&  \left [ \exp \left( \lambda_k e^{i \frac{k t}{\sigma \sqrt{\bar N}}}-
		\lambda_k-i \lambda_k \frac{k t}{\sigma \sqrt{\bar N}} \right ) \right]^{\bar N}.
\end{IEEEeqnarray}
The exponential in this expression can be taylor expanded as such that
\begin{equation}
	\chi_{z_k}(t)=\left[1-\frac{1}{2} \lambda_k \frac{k^2}{\sigma^2} \left( \frac{t^2}{\bar N} \right ) + \mathcal O \left( \frac{t^3}{\bar N^{3/2}} \right ) \right]^{\bar N}.
\end{equation}
In the limit $\bar N \to \infty$, the terms of order $t^3/ \bar N^{3/2}$ tend to zero. The second term is smaller than $1/2$, since $\lambda_k k^2<\sigma^2$. Thus, the well known asymptotic limit $\lim_{n \to \infty}(1-\frac{1}{2}a t^2/n)^n \longrightarrow \exp^{-\frac{1}{2}a t^2}$, can be used to show that $\chi_{z_k}(t)$ converge pointwise as
\begin{equation}
	\lim_{\bar N \to \infty} \chi_{z_k}(t) \longrightarrow e^{-\frac{1}{2} \lambda \frac{k^2 t^2}{\sigma^2}}.
\end{equation}
Subsequently,
\begin{equation}
	\lim_{\bar N \to \infty} \chi_Z(t) \longrightarrow e^{-\frac{1}{2}t^2}.
\end{equation}
By Levy's theorem, the discrete variable $Z(N)$ converges in distribution to the continuous variable $Z$. The characteristic function of which we recognise as that of the normal distribution with mean value zero and variance of one. Thus the probability density of $Z$ in the interval $[ Z,Z+\mathrm dz )$ is
\begin{equation}
	\rho(Z) \mathrm dZ = \frac{1}{\sqrt{2 \pi}} e^{-\frac{1}{2} Z^2} \mathrm dZ,
\end{equation}
On substitution of $(Z=N-\bar N)/\sigma_N$, we find that for large $\bar N$ the distribution $\gamma_N^2$ is well approximated by
\begin{equation} \label{eq:appgammaapprox}
	\gamma_N^2 \approx \frac{k_0}{\sqrt{2 \pi} \sigma_N} \exp \left[ - \frac{(N-\bar N)^2}{2 \sigma_N^2} \right],
\end{equation}
where $k_0$ is the greatest common denominator of $\{k\}$.

\end{document}